\documentclass{aa}

\usepackage{newtxtext,newtxmath}

\usepackage[T1]{fontenc}
\usepackage{ae,aecompl}

\usepackage{natbib}
\bibpunct{(}{)}{;}{a}{}{,} 

\usepackage{graphicx}	
\usepackage{amsmath}	
\usepackage{amssymb}	
\usepackage{xcolor}
\usepackage[normalem]{ulem}
\usepackage{subcaption}

\providecommand{\gaia}{\textit{Gaia }}
\providecommand{\gaianospace}{\textit{Gaia}}
\providecommand{\kms}{km s$^{-1}$ }

\providecommand{\msun}{$M_\odot$}
\providecommand{\lmc}{$\mathnormal{G}_{\text{LMC}}$ }
\providecommand{\smc}{$\mathnormal{G}_{\text{SMC}}$ }
\providecommand{\mw}{$\mathnormal{G}_{\text{MW}}$ }
\providecommand{\lmcnospace}{$\mathnormal{G}_{\text{LMC}}$}
\providecommand{\smcnospace}{$\mathnormal{G}_{\text{SMC}}$}
\providecommand{\mwnospace}{$\mathnormal{G}_{\text{MW}}$}

\definecolor{C0}{HTML}{1F77B4}
\definecolor{C1}{HTML}{FF7F0E}
\definecolor{C2}{HTML}{2CA02C}
\definecolor{C3}{HTML}{D62728}
\definecolor{C4}{HTML}{9467BD}
\definecolor{C5}{HTML}{8C564B}
\definecolor{C6}{HTML}{E377C2}
\definecolor{C7}{HTML}{7F7F7F}
\definecolor{C8}{HTML}{BCBD22}
\definecolor{C9}{HTML}{17BECF}
\definecolor{C10}{HTML}{CD853F}
\definecolor{C11}{HTML}{FF6347}

\newcommand{\kmskpc}{km s$^{-1}$ kpc$^{-1}$}

\begin{document}

\title{KRATOS: A large suite of N-body simulations \\ to interpret the stellar kinematics of LMC-like discs}

\author{Ó. Jiménez-Arranz\inst{1,2,3}
   \and S. Roca-Fàbrega\inst{4,5}
   \and M. Romero-Gómez\inst{1,2,3}
   \and X. Luri\inst{1,2,3}
   \and \\ M. Bernet\inst{1,2,3}
   \and P. J. McMillan\inst{6,4}
   \and L. Chemin\inst{7}
}

\institute{{Departament de Física Quàntica i Astrofísica (FQA), Universitat de Barcelona (UB), C Martí i Franquès, 1, 08028 Barcelona, Spain}
\and
{Institut de Ciències del Cosmos (ICCUB), Universitat de Barcelona, Martí i Franquès 1, 08028 Barcelona, Spain}
\and
{Institut d'Estudis Espacials de Catalunya (IEEC), c. Esteve Terradas 1, 08860 Castelldefels (Barcelona), Spain}
\and
{Lund Observatory, Division of Astrophysics, Lund University, Box 43, SE-221 00 Lund, Sweden}
\and
{Departamento de Física de la Tierra y Astrofísica, UCM, and IPARCOS, Facultad de Ciencias Físicas, Plaza Ciencias, 1, Madrid, E-28040, Spain}
\and
{School of Physics \& Astronomy, University of Leicester, University Road, Leicester, LE1 7RH, UK}
\and
{Instituto de Astrofísica, Universidad Andres Bello, Fernandez Concha 700, Las Condes, Santiago RM, Chile}
}

\date{Received <date> / Accepted <date>}

\abstract 
{The Large and Small Magellanic Cloud (LMC and SMC, respectively) are the brightest satellites of the Milky Way (MW) and, for the previous thousands of million years, they have been interacting with one another. Since observations only provide a static picture of the entire process, numerical simulations are used to interpret the present-day observational properties of these kinds of systems, and most of them have been focused on attempting to recreate the neutral gas distribution and characteristics through hydrodynamical simulations.}
{We present KRATOS, a comprehensive suite of 28 open access pure N-body simulations of isolated and interacting LMC-like galaxies, to study the formation of substructures in their disc after the interaction with an SMC-mass galaxy. The primary objective of this paper is to provide theoretical models that help interpreting the formation of general structures of an LMC-like galaxy under various tidal interaction scenarios. This is the first paper of a series that will be dedicated to the analysis of this complex interaction.} 
{Simulations are grouped in 11 sets of at most three configurations each containing: 1) a control model of an isolated LMC-like galaxy; 2) a model that contains the interaction with an SMC-mass galaxy, and; 3) the most realistic configuration where both an SMC-mass and MW-mass galaxies may interact with the LMC-like galaxy. In each simulation, we analyse the orbital history between the three galaxies and examine the morphological and kinematic features of the LMC-like disc galaxy throughout the interaction. This includes investigating the disc scale height and velocity maps. When a bar develops, our analysis involves characterising its strength, length, off-centeredness and pattern speed.} 
{The diverse outcomes found in the KRATOS simulations, including the presence of bars, warped discs, or various spiral arm shapes, demonstrate their capability to explore a range of LMC-like galaxy morphologies. Those directly correspond to distinct disc kinematic maps, making them well-suited for a first-order interpretation of the LMC's kinematic maps. From the simulations we note that tidal interactions can: boost the disc scale height; both destroy and create bars, and; naturally explain the off-center stellar bars. The bar length and pattern speed of long-lived bars are not appreciably altered by the interaction.}
{The high spatial, temporal, and mass resolution used in the KRATOS simulations has been shown to be appropriate for the purpose of interpreting the internal kinematics of LMC-like disks, as evidenced by the first scientific results presented in this work.}



\maketitle

\section{Introduction}
\label{sec:introduction}

Only a few extragalactic stellar structures are visible to the naked eye from Earth. The brightest of those are the Magellanic Clouds (MCs), which are the most massive of the Milky Way (MW) satellite galaxies. Because they are so close, the Large and Small Magellanic Cloud (LMC and SMC, respectively) provide astronomers with a unique window into the complexities of extragalactic systems. Furthermore, the evident large scale structures they contain, in particular the disc-like structure, the spiral arm and the bar in the LMC \citep[e.g.][]{Elmegreen1980,Gallagher1984,Yozin2014}, and the stellar bridge that connects both galaxies \citep[e.g.][]{harris07,Kallivayalil2013,Zivick2019,luri20}, make the LMC-SMC system an ideal laboratory to study the effects of galactic interactions on the evolution of dwarf galaxies and their structures with the current amount of data provided by the \gaia mission \citep{gaia2016b,gaiadr3}, VMC \citep{cioni11}, and SMASH \citep{Nidever2017}, to name a few. Additionally, more data is anticipated in the future thanks to projects like Euclid \citep{euclid12,euclid22}, the Vera C. Rubin Observatory \citep[previously referred to as LSST,][]{ivezic19}, SDSS-V \citep{kollmeier19,almeida23}, and 4MOST \citep{dejong19}.

The LMC is so peculiar that gives name to a type of galaxies, the Barred Magellanic Spirals \citep{vaucouleurs-freeman72}. This galaxy is a dwarf bulgeless spiral with a single spiral arm, an off-centred and asymmetric stellar bar, and many star forming regions \citep[e.g.][]{Elmegreen1980,Gallagher1984,Zaritsky2004,Yozin2014, luri20}. It is a gas-rich galaxy \citep[e.g.][]{luks-rohlfs92,kim98} characterised by an inclined disc \citep[e.g.][]{vdm01,vandermarel01}, with multiple warps reported \citep[e.g.][]{Olsen2002,nikolaev04,Choi2018,saroon22,ripepi22}, that lies at a distance of around 50 kpc \citep{pietrzynski19}. 
The SMC has long been thought to be a satellite of the LMC due to its proximity. It is at around 62 kpc from the MW \citep[e.g.][]{cionivdm00,hilditch05,graczyk14} and 20-25 kpc away from the LMC. The SMC is a gas rich dwarf irregular galaxy \citep[e.g.][]{rubio93,staveley-smith98}, and features a low metallicity environment \citep[e.g.][]{Choudhury18,grady21}. An attempt to reconstruct its 3D shape has been carried out using red clump stars and other standard candles and revealed that it is elongated about 15-30 kpc approximately in the east/north-east towards south-west direction \citep[e.g.][]{subramanian-subramanian12,ripepi17}.

Because of the lack of high-precision astrometric data it has been difficult to study the complex interaction between these two satellite galaxies and the MW. With the available data at the time, and by using theoretical models, many authors suggested that the MCs were in fact orbiting the MW, and that they had multiple experienced pericentric passages \citep[e.g.][]{tremaine76,murai80,lin82,gardiner94}. This conclusion was accepted for many years, till new data revealed a different history. The current accepted scenario was proposed by \citet{besla07} using data from HST \citep{kallivayalil06lmc,kallivayalil06smc} and theoretical models that showed that the MCs are most probably just after their first approach to the MW, with no prior pericenter passages within the last 10 Gyr. Moreover, the authors showed that the orbits of these two galactic systems around the MW are highly eccentric, with apocentres well beyond 200 kpc, and with orbital periods exceeding 5 Gyr. The debate on the date of the first pericenter is however not closed, for instance, \citet{vasiliev23b} claims that a scenario in which the LMC is on its second passage around the MW (that would have occurred 5-10 Gyr ago at a distance $\gtrsim$ 100 kpc) is consistent with current observational constraints on the mass distribution and relative velocity of both galaxies. 

Focusing now on the LMC-SMC system, there is also controversy on when the most recent interaction between the two satellites started. \citet{gardiner-noguchi96} stated that the two MCs most recent near encounter took place in between 150 and 200 Myr ago and that it was with a distance of less than 10 kpc. They also showed that this interaction would have led to the formation of the tidal structure known as Magellanic Bridge, which is a structure found in the region between the two galaxies \citep[e.g.][]{kerr54,misawa09}. First, \citet{hindman63} found that this tidal structure contains neutral hydrogen suggesting that it should have been formed recently and could host active star formation. Decades later, a stellar population of blue main-sequence stars in the Bridge was discovered by \citet{irwin85} confirming that star formation is ongoing within this structure. In the last years the Magellanic Bridge has been studied using both simulations \citep[e.g.][]{besla12,diaz-bekki12} and observations \citep[e.g.][]{harris07,Kallivayalil2013,bagheri13,noel13,skowron14,carrera17,Zivick2019,Schmidt20,luri20} trying to learn about its formation mechanism and also using it to better understand the complex interaction between the two satellite galaxies.

The study of the formation and evolution of the LMC-SMC system cannot be done only by using observational data. Observations give us only a static picture of the whole process, and that is why most researchers studying these kinds of systems use numerical simulations. These studies have naturally been focused on trying to recreate the distribution of neutral gas and the position and properties of the streams using N-body simulations that include hydrodynamics, with the goal of reproducing the whole past interaction process. 

For instance, \citet{pardy18} presented hydrodynamic simulations to reproduce the observations by \citet{hammer15} that showed that the Magellanic Stream is structured into two filaments \citep[e.g.][]{putman03b,nidever08}. In this study, the authors suggested that to reproduce the observations, the MCs should have been more gas-rich in the past, and that the gas stripping efficiency of the LMC should have been much higher. Later, \citet{wang19} show, for the first time, that a physical modelling is capable to explain and reproduce the enormous quantities of gas stripped from the MCs, namely more than 50 per cent of their initial content. More recently, \cite{tepper-garcia19} include for the first time a weakly magnetised and spinning Magellanic Corona, a halo of warm gas surrounding the LMC and SMC, in their simulations to reproduce the location and the extension of the Magellanic Stream on the sky. \cite{lucchini20} also included a Magellanic Corona to show that its presence can explain the ionised gas component of the Magellanic Stream. Finally, \citet{lucchini21} present new simulations of the formation of the Magellanic Stream with a new first-passage interaction history of the MCs where the orientation of the SMC’s orbit around the LMC is qualitatively different and leads to a different 3D spatial positioning of the Stream from previous models. Their simulated Stream is only at $\sim$20 kpc away from the Sun at its closest point, whereas previous first-infall models predicted a distance of 100-200 kpc.

Similarly, the study of the interaction between the MCs and the MW can not only rely on observations but requires again  the use of numerical simulations. As mentioned before, several N-body simulations have been run and used to analyse the effect of the MCs (or, more specifically, the LMC) on the MW \citep[e.g.][]{garavito-camargo19,petersen-peñarrubia22} and to determine the MCs past orbits \citep[e.g.][]{vasiliev23,vasiliev23b}. However, the study of the effect of these interactions on the internal structures of the LMC-like disc, the bar and the spiral arm has only been carried out by a few authors. In particular, only the work by \citet{besla12} extensively explored these features in simulations. Understanding the formation process of these LMC morphological attributes can potentially unveil the details on the interaction occurred between these two satellite galaxies, and also of them with the MW.

In this context, here we present KRATOS, a comprehensive suite of 28 open access pure N-body simulations of isolated and interacting LMC-like and SMC-mass galaxies. With these models we study the formation of stellar substructures in an LMC-like disc after the interaction with an SMC-mass system and we compare them with the observations \citep[see, for example, the kinematic maps of the LMC using \gaia data on][]{luri20,jimenez-arranz23a}. This is the first paper of a series that will be dedicated to the analysis of this complex interaction. In this work, we show the high degree of detail of the simulations and, as a first scientific case, we study the orbital history between the three galaxies and the evolution of the LMC-like morphological and kinematic features, such as the kinematic maps, the disc scale height and the properties of the bar, when this is formed. A more specific analysis on the LMC-SMC interaction is left for successive papers of the series.

The paper is organised as follows. In Sect. \ref{sec:description}, we describe the code, initial conditions and tools used to run and analyse the KRATOS suite. In Sect. \ref{sec:orbital_history}, we characterise the orbital history between the three galaxies. In Sect. \ref{sec:lmc_properties}, we study the morphological and kinematic features of the LMC-like galaxy at present time. In Sect. \ref{sec:scale_height}, we discuss how the scale height of the LMC-like disc changes with the different pericenters of the SMC-mass system. In Sect. \ref{sec:lmc_bar}, we study the properties' evolution of the LMC-like galaxy bar. In Sect. \ref{sec:discussion}, we contextualise our results with the LMC observations and other works in the literature. Finally, in Sect. \ref{sec:conclusions}, we summarise the main conclusions of this work.

\section{KRATOS simulations}
\label{sec:description}

KRATOS (Kinematic Reconstruction of the mAgellanic sysTem within the OCRE Scenario) consists of 28 pure N-body simulations of isolated and interacting LMC-like and SMC-mass galaxies\footnote{The simulations are open access. Readers interested in using the simulations developed in the paper can access them at \url{https://dataverse.csuc.cat/dataset.xhtml?persistentId=doi:10.34810/data1156}.}. In the context of galactic simulations, the suffix "-like" is designated to a system that is similar in both mass and shape, while the suffix "-mass" is designated to a system that is only similar in mass. In these models, we do not include hydrodynamics or cosmological environment. The 28 models are grouped in 11 sets of at most three models each containing: 1) a control model with an isolated LMC-like galactic system; 2) a model with both an LMC-like and a SMC-mass system; 3) a model that additionally includes a MW-mass system. Hereafter we refer to the LMC-like, SMC-mass, and MW-mass galactic systems as \lmcnospace, \smcnospace, and \mwnospace, respectively.  By implementing these three scenarios, we aim to distinguish the local instabilities in the \lmc disc from the products of the interactions between these galaxies. For each of the three scenarios, we vary a set of free parameters, namely, the \lmc disc instability (given by the Toomre $Q$ parameter), the \lmc disc mass, the \lmc halo mass, the \smc mass, the \mw mass and the \mw halo mass distribution. In order to better understand the effect of each of the free parameters on the \lmcnospace-\smc interaction and on the formation of \lmcnospace's disc structures, we vary one parameter at a time.

Each model in the KRATOS suite has been run within a $2.85^3$ Mpc$^3$ box with periodic boundary conditions. The simulations have a spatial and temporal resolution of 10 pc and 5000 yr, respectively. The minimum mass per particle is $4 \times 10^3$\msun. All simulations have been run for 4.68Gyr, starting at the apocenter between the MCs after their second interaction. According to \citet{lucchini21}, this happened 3.5 Gyr ago. Thus, we have run the simulations for more than one gigayear after the MCs match the most recent observations (see Sec.~\ref{subsec:initial_conditions}).

\subsection{The code}
\label{subsec:the_code}

The numerical simulations have been computed using the Eulerian pure N-body code {\sc ART} \citep{Kravtsov97}. The code is based on the adaptive mesh refinement technique, which allows to selectively boost resolution in a designated region of interest surrounding a chosen dark matter (DM) halo.

Most of the KRATOS suite was run in virtual machines in the Google cloud provided by the Open Clouds for Research Environments (OCRE) project funded by the European Union’s Horizon 2020 research and innovation program. We used 24 virtual machines, each with 16 cores, 128GB RAM, and 250GB SSD to run each simulation independently (and simultaneously) for three weeks. Four additional simulations have been run in the Brigit supercomputer of the Universidad Complutense de Madrid, using 16 cores each. The full set of simulations amount to a total of 285~000 hours of computational time.

\subsection{Initial conditions}
\label{subsec:initial_conditions}

As mentioned earlier, our approach involves systematically varying the parameters individually, one per simulation. The initial conditions for the construction of the fiducial \lmcnospace, \smcnospace, and \mw galaxies are outlined in Table \ref{tabl:kratos_simulations_fiducial}, whereas the initial conditions for the other simulations of the suite are outlined in Table  \ref{tabl:kratos_simulations_all}. The colour code used for each set is kept all throughout the paper. The initials conditions  were produced using the RODIN code as in \citet{roca-fabrega12,roca-fabrega13,RocaFabrega2014}.

\begingroup

\setlength{\tabcolsep}{10pt} 
\renewcommand{\arraystretch}{1.5} 
\begin{table*}
\centering
\begin{tabular}{llll}
\hline
         &  \lmc  & \smc  &  \mw \\
\hline
DM Concentration   & 9 & 15 & 12 \\
DM Mass $(M_\odot)$   & $1.8 \times 10^{11}$  & $1.9 \times 10^{10}$  &  $10^{12}$   \\

Stellar Mass $(M_\odot)$    & $5.0 \times 10^{9}$ & $2.6 \times 10^8$ & -      \\
Toomre parameter $Q$  & 1.2 & - & -      \\
Stellar Scale Height (kpc)  & 0.20 & - &  -   \\
Stellar Scale Length (kpc)     & 2.85 & - & -  \\
Stellar Disc Truncation Radius (kpc)   & 11.5 & - & -    \\
Number of Stellar Particles   &  1~200~002  & 62~402 & -    \\
Number of DM Particles    & 24~281~795   & 4~497~354 & 1~999~733 \\
DM Species    & 7 & 1 & 1 \\
Mass Resolution $(M_\odot)$   & $4\times10^3$ & $4\times10^3$  & $5\times10^5$ \\
\hline
Initial Position (kpc)  & (0,\hspace{0.1cm}0,\hspace{0.1cm}0) & (-67.15, -134.09, 33.23) &  (-47.36, -546.38, -150.52) \\
Initial Velocity (km/s)     & (0,\hspace{0.1cm}0,\hspace{0.1cm}0) & (11.72, 21.81, -16.60) & (-1.71, 99.02, 63.73) \\
\hline
\end{tabular}
\caption{Initial conditions of the fiducial model presented in this work.}
\label{tabl:kratos_simulations_fiducial}
\end{table*}

\endgroup

\begingroup

\setlength{\tabcolsep}{10pt} 
\renewcommand{\arraystretch}{1.5} 
\begin{table*}
\centering
\begin{tabular}{llll}
\hline
         &  Model label  & Configuration  &  Changes of each set with respect to the fiducial model \\
\hline 
\color{C0}{\textbf{K1}}   & Fiducial & \lmc & -  \\
\color{C0}{\textbf{K2}}   & (blue) & \lmcnospace+\smc &  \\
\color{C0}{\textbf{K3}}   &  & \lmcnospace+\smcnospace+\mw &  \\
\hline 
\color{C10}{\textbf{K4}}   & $Q=1.0$ & \lmc & [\lmcnospace] Toomre parameter $Q=1.0$  \\
\color{C10}{\textbf{K5}}   & (light brown) & \lmcnospace+\smc &  \\
\color{C10}{\textbf{K6}}   &  & \lmcnospace+\smcnospace+\mw &  \\
\hline 
\color{C11}{\textbf{K7}}   & $Q=1.5$ & \lmc & [\lmcnospace] Toomre parameter $Q=1.5$  \\
\color{C11}{\textbf{K8}}   & (light red) & \lmcnospace+\smc &  \\
\color{C11}{\textbf{K9}}   &  & \lmcnospace+\smcnospace+\mw &  \\
\hline 
\color{C1}{\textbf{K10}}   & $0.45\text{M}_{\text{LMC}}^{\text{halo}}$ & \lmc & [\lmcnospace] DM Mass $= 0.8 \times 10^{11} M_\odot$  \\
\color{C1}{\textbf{K11}}   & (orange) & \lmcnospace+\smc & \hspace{1.1cm} Number of DM Particles = 11 992 066   \\
\color{C1}{\textbf{K12}}   &  & \lmcnospace+\smcnospace+\mw &  \\
\hline 
\color{C2}{\textbf{K13}}   & $1.40\text{M}_{\text{LMC}}^{\text{halo}}$ & \lmc & [\lmcnospace] DM Mass $= 2.5 \times 10^{11} M_\odot$  \\
\color{C2}{\textbf{K14}}   & (green) & \lmcnospace+\smc &  \hspace{1.1cm} Number of DM Particles = 34 924 737   \\
\color{C2}{\textbf{K15}}   & & \lmcnospace+\smcnospace+\mw &   \\
\hline 
\color{C7}{\textbf{K16}}   & $0.60\text{M}_{\text{LMC}}^{\text{disc}}$  & \lmc & [\lmcnospace] Stellar Mass $= 3.0 \times 10^{9} M_\odot$ \\
\color{C7}{\textbf{K17}}   & (grey)  & \lmcnospace+\smc & \hspace{1.1cm} Number of Stellar Particles = 720 005  \\
\color{C7}{\textbf{K18}}   &  & \lmcnospace+\smcnospace+\mw &  \\
\hline 
\color{C8}{\textbf{K19}}   & $0.60\text{M}_{\text{LMC}}^{\text{disc}}$; ${ } 0.45\text{M}_{\text{LMC}}^{\text{halo}}$  & \lmc & [\lmcnospace] Stellar Mass $= 3.0 \times 10^{9} M_\odot$ \\
\color{C8}{\textbf{K20}}   & (yellow) & \lmcnospace+\smc & \hspace{1.1cm} Number of Stellar Particles = 720 005   \\
\color{C8}{\textbf{K21}}   & & \lmcnospace+\smcnospace+\mw & [\lmcnospace] DM Mass $= 0.8 \times 10^{11} M_\odot$    \\
 &  & & \hspace{1.1cm} Number of DM Particles = 11 992 066  \\
\hline 
\color{C9}{\textbf{K22}}   & $0.60\text{M}_{\text{LMC}}^{\text{disc}}$; ${ } 1.40\text{M}_{\text{LMC}}^{\text{halo}}$ & \lmc & [\lmcnospace] Stellar Mass $= 3.0 \times 10^{9} M_\odot$  \\
\color{C9}{\textbf{K23}}   & (cyan)  & \lmcnospace+\smc & \hspace{1.1cm} Number of Stellar Particles = 720 005  \\
\color{C9}{\textbf{K24}}   &  & \lmcnospace+\smcnospace+\mw  & [\lmcnospace] DM Mass $= 2.5 \times 10^{11} M_\odot$   \\
 &  & & \hspace{1.1cm} Number of DM Particles = 34 924 737  \\
\hline 
\color{C3}{\textbf{K25}}   & $0.25\text{M}_{\text{SMC}}$ & \lmcnospace+\smc & [\smcnospace] DM Mass $= 0.5 \times 10^{10} M_\odot$  \\
\color{C3}{\textbf{K26}}   & (red) & \lmcnospace+\smcnospace+\mw & \hspace{1.1cm} Number of DM Particles = 1 199 900   \\
\hline 
\color{C5}{\textbf{K27}}   & $0.15\text{M}_{\text{MW}}$ & \lmcnospace+\smcnospace+\mw &  [\mwnospace] DM Mass $= 0.15 \times 10^{12} M_\odot$  \\
 & (dark brown) &  &    \\
\hline 
\color{C6}{\textbf{K28}}   & Point-like MW & \lmcnospace+\smcnospace+\mw  & [\mwnospace] Point-like \\
 & (pink) &   &  \\
\hline 
\end{tabular}
\caption{Initial conditions of all the simulations presented in this work with respect to the fiducial model. The color code used for each set is kept all throughout the paper.}
\label{tabl:kratos_simulations_all}
\end{table*}

\endgroup

In all simulations, we model the \lmc system as a stellar exponential disc embedded in a live dark matter Navarro-Frenk-White \citep[NFW, ][]{nfw96} halo. Its stellar disc has a scale length and scale height of 2.85 and 0.20 kpc, respectively. The initial disc scale length is higher than the one used in similar works \citep[e.g.][]{besla12,lucchini21}, but as the disc relaxes, it decreases to 1-2 kpc, which is consistent with estimates provided by \citet{fathi10} for galaxies with low stellar mass. We consider a \lmc disc truncation radius of 11.5 kpc. The \lmc NFW DM halo has a concentration of $C=9$ \citep{besla10,besla12,lucchini21}. Its DM halo is composed of 7 species of DM particles, each with twice the mass of the previous one, with the most massive ones being the farthest from the disc. It has been shown that the contamination by massive dark matter particles in the region of the disc is low, as discussed in \citet{ValenzuelaKlypin03}.

The \smc system is modelled as a simple NFW halo with a concentration of $C=15$ \citep{besla10,besla12,lucchini21}. Both \smc dark matter and stellar particles are generated at once following the NFW profile. For visualisation and analysis purposes we later define the stellar component of the \smc as the particles that have the strongest gravitational binding. This selection was carried out until the cumulative mass of the chosen particles equaled the baryonic matter mass observed in the SMC. To ensure that the inner region is not depleted of DM particles, we select one out of every two particles as a star particle. This selection process does not have any impact on the models as all particles, both DM and stellar, are treated as collisionless point-like sources of gravity. By employing this particle selection strategy, we just aimed to capture the evolution of the stellar component and its interaction with the surrounding environment. We do not include a stellar disc for the \smc because old and intermediate-age stars are distributed in a spheroidal or slightly ellipsoidal component \citep[the SMC is an irregular galaxy, e.g.][]{subramanian-subramanian12}, and because we are interested in the gravitational interaction between \lmc and \smcnospace.

Finally, since we are mostly interested in the effects that the interaction between the three galaxies produces in the \lmc disc, we only model the MW DM content in \mwnospace, thus neglecting the contribution of the MW disc to the total mass of \mwnospace. We employ an NFW profile whose concentration parameter chosen was set to $C=12$ \citep{besla10,besla12,lucchini21}. All three objects are present at the beginning of the simulation.

The fiducial simulation has the same initial conditions as the simulations performed in \citet{lucchini21} for the density, kinematics and orbital parameters. The main differences between their work and our simulations are: 1) all models in the KRATOS suite are pure N-body whereas their simulation considers hydrodynamics, and; 2) the KRATOS suite has a higher spatial, temporal, and mass resolution. We consider a \lmc disc with a mass of $5 \times 10^9$\msun, a high-mass disc if we compare it with observations \citep[e.g.][]{vandermarel02}. The \lmc disc Toomre $Q$ parameter is 1.2. The \lmc system has a total DM mass of $1.8 \times 10^{11}$\msun. We consider a \smc system with DM mass of $1.9 \times 10^{10}$\msun, and baryonic mass of $2.6 \times 10^8$\msun. The DM mass of the \mw system DM is considered to be $10^{12}$\msun.  Finally, regarding the orbital parameters, we chose as the starting point the \smc being at the second apocenter of the LMC-SMC interaction, which occurred 3.5 Gyr ago. We choose this approximation because in the present time most morphological and kinematic footprints of this very past interaction would have been already erased by other internal and external processes within each one of the systems. For our fiducial model we also set the orbit of the \smc around the \lmc as being prograde. Table \ref{tabl:kratos_simulations_fiducial} summarises the initial conditions of the fiducial model.

The variations of the different parameters with respect to the fiducial simulation that we consider are: 1) a lighter \lmc disc with a baryonic mass of $3 \times 10^9$\msun; 2) a lighter and a heavier \lmc DM halo with a mass of $0.8 \times 10^{11}$\msun~and $2.5 \times 10^{11}$\msun, respectively; 3) a lighter \smc with DM mass of $0.5 \times 10^{10}$\msun; 4) a \mw system almost an order of magnitude lighter, with mass equal to $0.15 \times 10^{12}$\msun, to also cover the lowest \mw estimations \citep{jiao23}; 5) a \mw system modelled as a single particle of $10^{12}$ \msun~(point-like mass approximation), to test the effect on the \lmcnospace-\smc interaction of changing the DM distribution. Table \ref{tabl:kratos_simulations_all} summarises the differences between the 11 sets of simulations.

The absence of hydrodynamics in our simulations implies some limitations, particularly in understanding the gaseous features of the MCs, such as the Leading Arm, the Magellanic Stream, and the gaseous component of the Magellanic Bridge. Additionally, the lack of hydrodynamic modelling precludes an exploration of stellar evolution resulting from interactions between the MCs. Also, the exclusion of the gas limits our insights into the self-sustainability of the arms by means of an intrinsically colder disc. These limitations and others should be kept in mind when interpreting the results and highlight potential directions for future studies incorporating hydrodynamic simulations.

\subsection{Centering, alignment and bar's pattern speed}
\label{subsec:analysis_tools}

To study the general features of the \lmc disc we take the center-of-mass of the \lmc baryonic matter as the center of the reference frame (see Sect.~\ref{sec:lmc_properties}). On the other hand, in Sect. \ref{sec:lmc_bar}, we are interested in studying the \lmcnospace's bar, which is not located in the center-of-mass of the system when tidally perturbed, thus, we redefine the reference frame center to the one defined by the bar's density center. To find its density center, we first sample each particle's coordinate $(x,y,z)$ in the range -6.2 kpc to 6.2 kpc with 301 equidistributed bins. Later, we apply a Gaussian kernel density estimation (KDE) of 3.0 kpc-bandwidth and we take the point with the highest KDE density as the reference center. This approach involved testing different bandwidth values to identify and select the most suitable value for determining the bar center.

Defining the galactic disc plane is also a difficult task, especially when the \smc and \mw interact with the \lmcnospace. In these interactions, the \lmc disc suffers strong perturbations sometimes almost destroying the disc. In this situation, the strategy to define the \lmc galactic disc plane is to compute the angular velocity vector $\Vec{L}$ of all the disc's stars and take the perpendicular plane.

Finally, in Sect. \ref{sec:lmc_bar}, we use the Dehnen method \citep{dehnen23}, which was also used to determine the LMC bar pattern using \gaia DR3 data in \citet{jimenez-arranz23c}, to find the pattern speed of \lmc bars generated in our simulations. This method measures the bar pattern speed $\Omega_p$ and the orientation angle $\phi_b$ of the bar from single snapshots of simulated barred galaxies. For more details about the method see Sect~2 and Appendix~B of \citet{dehnen23}.

\section{The MCs orbital history}
\label{sec:orbital_history}

In this section, we study the orbital history of the three galaxies. We notice that our results differ from the one by \citet{lucchini21} in the time of the closest approach to the \mw system. Whereas \citet{lucchini21} needed to run the simulation for 3.46 Gyr to obtain two pericenter passages between the \lmc and \smc galaxies (see their Figure 1 right panel), we needed almost half a gigayear more for the fiducial simulation (K3). The origin of this discrepancy is still unclear but, as mentioned above, the main differences between our models and the ones in \citet{lucchini21} are their lower spatial and mass resolution, and the lack of the hydrodynamical content in ours. The almost one order of magnitude on the spatial resolution can drive big differences on the interaction times due to how accurately the individual orbits of stars and dark matter are calculated (see e.g. Roca-Fàbrega et al. 2023, submitted to ApJ). Also, baryonic processes like SNe feedback can modify the density distribution of the central halo which would lead to a change on the acceleration suffered by the \smc system and, thus, on the interaction times \citep[e.g.][]{Duffy2010}. The Lucchini model has a hot gas corona of $10^{11}$\msun, which has roughly the same effect on the gravitational forces on the MCs as giving the dark-matter halo an extra $10^{11}$\msun${ }$ in mass. Nevertheless, our examination of the effects of incorporating this additional mass into the \mw model reveals no significant variations. Finally, they use a Hernquist halo for the dark matter, not a NFW, which means the mass is more centrally concentrated, i.e., having something like the effect of having a point-like MW (though less extreme). We chose the NFW profile over the Hernquist profile based on the consideration that the NFW distribution provides a better fit to the distribution of dark matter halos in cosmological simulations \citep[e.g.][and references therein]{lilley18}.

In this scenario, as the interaction history of our fiducial simulated galactic system (K3) do not match with the ones obtained in \citet{lucchini21}, we need to determine at which time our simulated \lmcnospace+\smcnospace+\mw system most resembles the LMC observations, i.e. to set what we call the "present time" ($t=0$) for further analysis. We decide to set as the $t=0$ the snapshot after 4.0 Gyr from the initial conditions for the two following observational constraints: 1) if we assume that the \lmcnospace+\smc system already went through two pericenters, our fiducial simulation needs to evolve, at least, for 4.0 Gyr (see Sect. \ref{subsec:orbit_SMC_LMC}); 2) the real observed LMC disc morphology has very characteristic features such as a single spiral arm and an off-center bar, and these features are observed in the \lmc disc of the \lmcnospace+\smcnospace+\mw fiducial simulation just at this time (see Sect. \ref{sec:lmc_properties}).

Setting $t=0$ to be 4.0 Gyr after the initial conditions has the consequence that the distance between the \lmc and \mw (see Section \ref{fig:distance_MW}) is significantly larger ($\sim$ 200 kpc) than the observations \citep[$\sim$ 50 kpc][]{pietrzynski19}. However, if we let the system evolve until the distance between the \lmc and \mw is $\sim$ 50 kpc we would have three \lmcnospace-\smc pericenters (see Section \ref{fig:distance_SMC}). After careful consideration, we opted to keep the constraint of the two pericenters scenario recognising the deviation from observational data in a noticeably larger \lmcnospace-\smc distance and acknowledging the implications for the system's evolution.

\subsection{\smc to \lmc distance}
\label{subsec:orbit_SMC_LMC}

\begin{figure}[t!]
    \centering
    \includegraphics[width=0.87\columnwidth]{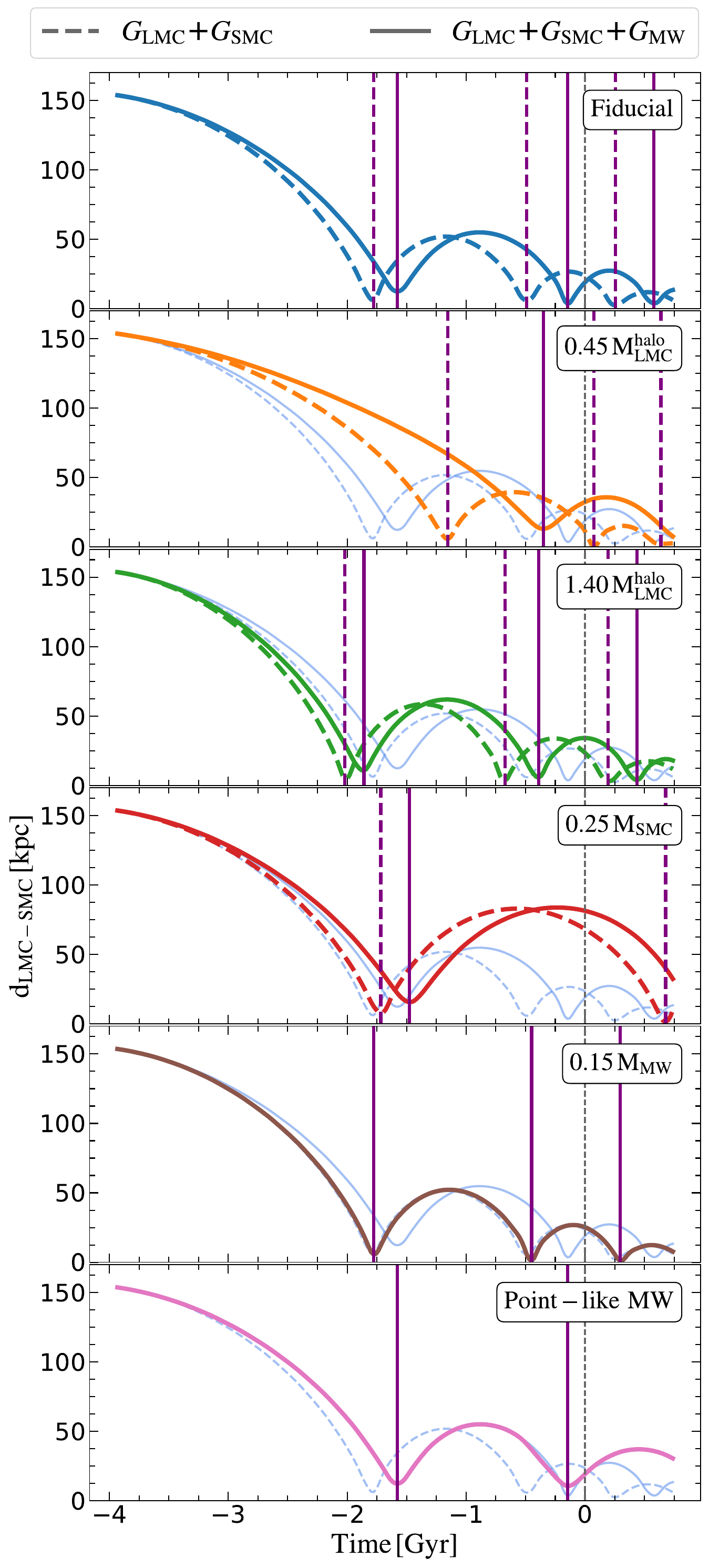}
    \caption{Distance between the centre-of-masses of the \lmc and \smc galaxies. The solid (dashed) lines corresponds to the \lmcnospace+\smcnospace+\mw (\lmcnospace+\smcnospace) models. The different panels represent a different set of simulations (see labels within each panel). The vertical purple solid (dashed) lines correspond to the \lmcnospace-\smc pericenters of the \lmcnospace+\smcnospace+\mw (\lmcnospace+\smcnospace) models. The vertical grey dashed line corresponds to the time we choose as the present time $t=0$ (see Sec.~\ref{subsec:initial_conditions}). For the sake of comparison, in shadowed blue lines it is plotted the orbital history of the MCs for the models of the fiducial set. The set corresponding to a different \lmc disc mass and Toomre parameter are not shown since they show no difference in comparison to the fiducial set. }
    \label{fig:distance_SMC}
\end{figure}

In Fig. \ref{fig:distance_SMC} we show the distance between the centre-of-mass of the \lmc and the \smc galaxies as a function of time. In solid lines we show the \lmcnospace+\smcnospace+\mw models whereas in dashed lines we show the \lmcnospace+\smcnospace. Purple vertical solid and dashed lines indicate the times of the pericenters, for each one of the models.

In the top panel, we show results from the fiducial simulations (blue lines), which are then used as guiding lines in all other panels (thin shadowed blue lines). The fiducial model shows that the time of the pericenters changes when including the gravitational pull of a \mw system. In particular, when the \mw system is present (solid lines), the pericenter is delayed for about 200 Myr. This result is a consequence of that the \smc is initially located in between the \mw and the \lmc galaxies, so the gravitational pull partially cancels-out, thus, the \smc infall towards the \lmc galaxy is slower and delayed. This same effect can also be observed in all other models (see solid vs. dashed lines in all other panels). The presence of a \mw system also has an impact on the orbit of the \smc around the \lmc galaxy. We can see that the minimum distance at pericenter is also different between the two models (with and without the \mw system). While at the first pericenter the distance is smaller when the \mw system is not present (6.2 kpc vs. 12.3 kpc), in the second the situation is reversed. This is a clear sign that the orbit has been slightly modified, i.e., that the \smc system has a different energy and angular momentum due to the gravitational pull of the \mw system. Although not analysed in this figure, this difference of the minimum distance also has a strong impact on the morphology of the \lmc galaxy disc (see Fig. \ref{fig:macro_density} and Sect. \ref{sec:lmc_properties}), with the disc of the \lmcnospace+\smc models initially more perturbed than the one in the \lmcnospace+\smcnospace+\mwnospace. 

In the second and third panels, we show the effect of changing the mass of the \lmc galaxy DM halo. First, we show that a lighter DM halo (orange lines) produces a delay on the pericenters, while a heavier DM halo (green lines) has the opposite effect. These differences have the same origin as the ones between the \lmcnospace+\smcnospace+\mw and the \lmcnospace+\smc models (solid vs. dashed lines, respectively), that is a change on the resulting acceleration over the \smc galaxy by the \mw and the \lmc galaxies. Like in the first panel, the variation on the pericenter time produces a change on the orbit of the \smc system bringing it closer to the \lmc galaxy in the first pericenter when it happens earlier, i.e. the \smc system is in a more radial orbit when the gravity of the \lmc galaxy dominates the interaction. 

In the fourth panel, we show the effect of reducing the \smc system mass (red line). We see that while the first pericenter does not change significantly from the one of the fiducial model (blue solid and dashed shadowed lines), there is a strong divergence afterwards. In particular, we see that the orbit of the \smc system decays slower when it is lighter. This is not surprising as it is well known that the dynamical friction is more efficient for high-mass than for low-mass systems \citep[e.g. Sect. 8.1 of][]{binney-tremaine08,chandrasekhar43}.

The effect of changing the total mass and the mass distribution of the \mw system is shown in the fifth and sixth panels (dark brown and pink lines), respectively, and is similar, but with the opposite sign, to changing the mass of the \lmc galaxy DM halo. A lighter \mw system makes the gravity of the \lmc galaxy dominate the \smc system's infall, i.e. the first pericenter occurs earlier and the orbit is more radial (smaller \lmcnospace-\smc distance at pericenter), similar to the \lmcnospace+\smc fiducial model where \mw is not present. On the other hand, keeping the \mw system's mass but changing its distribution to a much less realistic point-mass approximation has two effects. First, the \mw system has the same effect as if it was more massive, i.e. the pericenter is delayed with respect to the models with a lighter \mw system. This is because in the model with a \mw system with a mass distributed in a NFW profile the \lmcnospace-\smc system falls into the \mw DM halo much before the interaction between them starts, thus, a non-negligible fraction of the \mw system's mass do not contribute to the total acceleration applied to the \smc system (i.e. Gauss theorem). This can also be deduced by the fact that the pink solid line perfectly overlaps the blue shadowed line (fiducial \lmcnospace+\smcnospace+\mw model) almost down to the second pericenter when the \mw system has a close encounter with the \lmcnospace+\smc system, then the model significantly diverges from the fiducial. We note that considering a point-like MW also changes the dynamical friction of the interaction. Secondly, the model with a point-like \mw system all mass is pulling the \smc in a single direction, all the time, this is the origin of the second effect we observe that is a big difference on the \smc system's orbit. The \smc system experiences a strong and well directed tide by the \mw system, specially strong when it gets closer. As a consequence, the \smc total energy-angular momentum and, thus, its orbit, highly differs from the one in other models. This can be seen by comparing the behaviour of the pink solid line just before the second pericenter and later, when the \smc system follows a more circular orbit (larger radii and longer period between pericenters). 

Finally, notice that in this figure we do not show the results from the set of models with a smaller \lmc disc mass and different Toomre parameter since the total mass of the system is the same and, thus, they show no difference in the \lmcnospace-\smc orbital analysis with respect to the fiducial set.

\subsection{\lmcnospace-\smc system to \mw distance}
\label{subsec:orbit_MCs_MW}

\begin{figure}[t!]
    \centering
    \includegraphics[width=0.9\columnwidth]{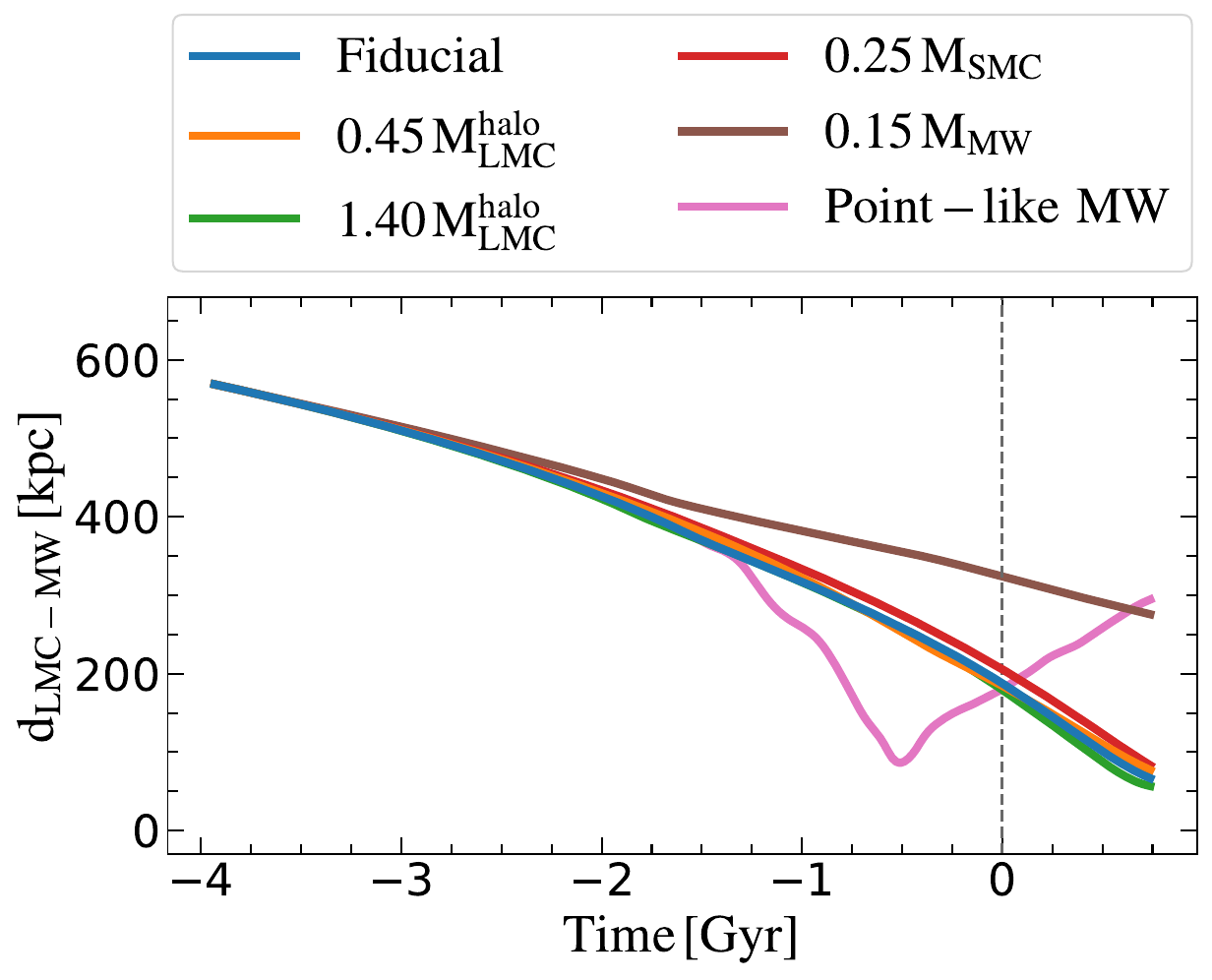}
    \caption{Distance between the centre-of-masses of the \lmc and \mw galaxies. Each color represents the \lmcnospace+\smcnospace+\mw model of a different set. The vertical grey dashed line corresponds to the present time $t=0$.}
    \label{fig:distance_MW}
\end{figure}

Figure \ref{fig:distance_MW} shows the distance between the centre-of-mass of the \lmc and \mw galaxies. We show that, overall, there is no big difference between models except for two, the model with a smaller \mw galaxy mass (dark brown line) and the point-like \mw model (pink line). In the former, the mutual acceleration between the \lmc galaxy and the \mw system is smaller, so they approach each other slower (dark brown line). In the latter, although initially similar (an extended object interacts as a single-point mass when far enough), the interaction becomes stronger when the \mw system approaches the \lmc galaxy and reaches pericenter earlier ($\sim$-0.5 Gyr), that is also when we observe big differences with the fiducial \lmcnospace-\smc distance (see Sec.~\ref{subsec:orbit_SMC_LMC}). The variations on the \lmcnospace-\mw distance when changing the \lmc and \smc masses are negligible as it is the \mw system's mass that dominates the dynamics of the \lmc-\mw interaction. Notice that, as in the previous section, we do not analyse the set of models where we changed the \lmc disc mass and Toomre parameter as these are almost identical, in total mass, to the fiducial set.

We also see that for the fiducial model (and most of the other models) the \mw is at a $\sim$200 kpc distance from the \lmc at the present time $t=0$. This distance is much larger than the one obtained from observations \citep[$\sim$50 kpc,][]{pietrzynski19}. This is a result that differs from the results presented by \citet{lucchini21}, and its origin can be in the difference in spatial and mass resolution between our models, as discussed above. For completeness, we also ran the \lmcnospace+\smcnospace+\mw fiducial model for one extra Gyr to find when the \mw system gets as close to the \lmc galaxy as in the observations (i.e. 50 kpc). The result is that this happens only after the \lmcnospace-\smc went through three pericenters instead of the two predicted by \citet{lucchini21}. Thus, we keep $t=0$ as the snapshot after 4.0 Gyr from the initial conditions for the reasons given above.

\section{The \lmc galaxy properties}
\label{sec:lmc_properties}

\subsection{t=0 morphologies}

\begin{figure*}[t!]
    \centering
    \includegraphics[width=0.75\paperwidth]{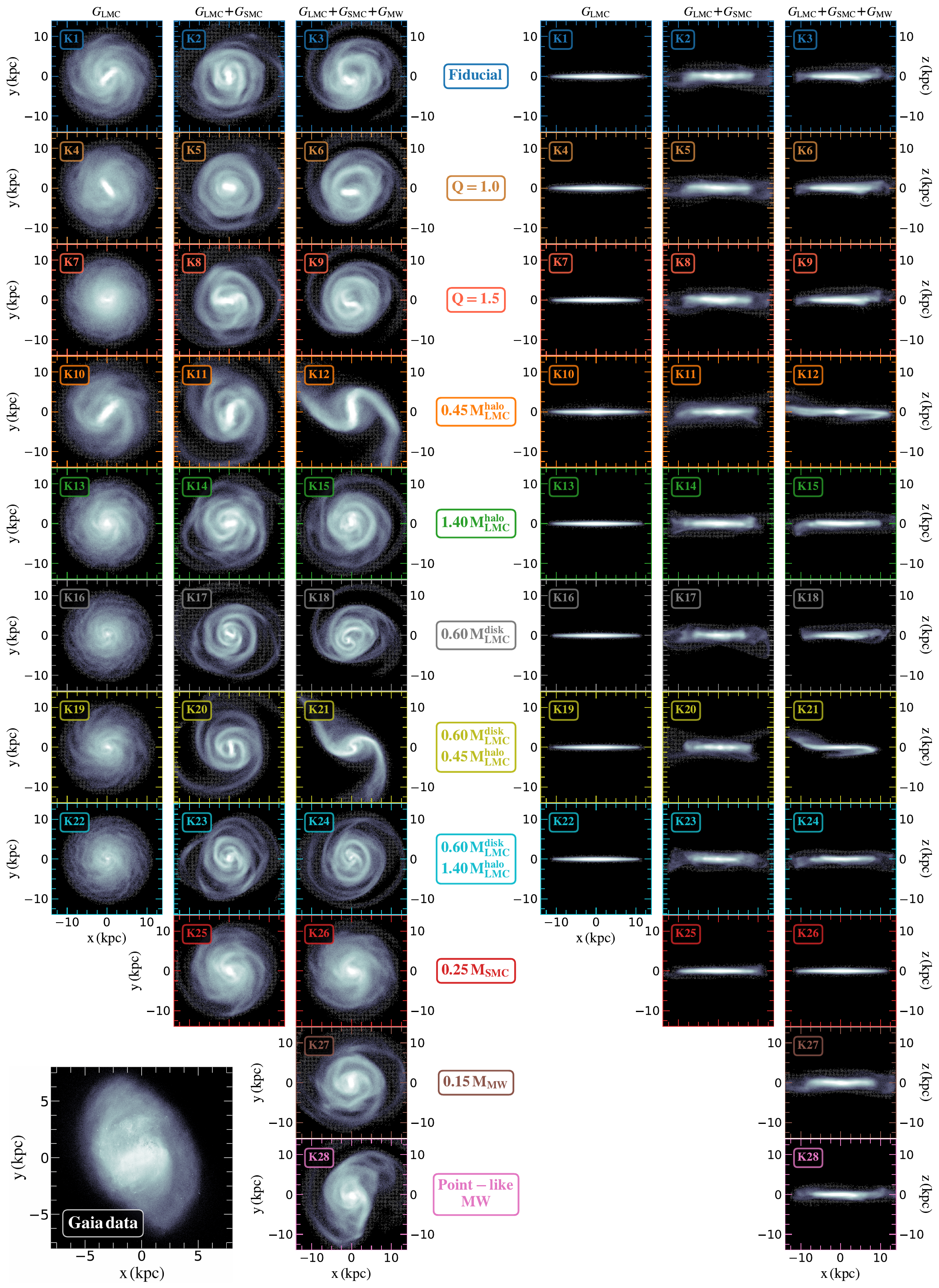}
    \caption{Stellar density map of the \lmc disc as seen face-on (left part) and edge-on (right part) at $t=0$. Each row corresponds to a different set of simulations and the labels are displayed in the rightmost panels. For both face-on and edge-on representations, we have the \lmcnospace, \lmcnospace+\smc and \lmcnospace+\smcnospace+\mw models on the left, centre and right panels, respectively. As reference, the face-on density map of the LMC optimal sample \citep{jimenez-arranz23a} is shown in the bottom left part.}
    \label{fig:macro_density}
\end{figure*}

In Fig. \ref{fig:macro_density} we show the face-on (left columns) and edge-on (right columns) of the \lmc disc stellar density at $t=0$ for all models\footnote{The animation showcasing the evolutionary changes in the face-on and edge-on distributions throughout the entire simulation is available online.}. Each row shows a different set of models (see legend in the central panels of each row). All \lmc galactic discs have been centred in the centre-of-mass and aligned following the procedure described in Sect. \ref{subsec:analysis_tools}. We warn the reader that, even though it is the same instant in time for all simulations, we may not be looking at the same stage of the \lmcnospace-\smc interaction (see Figs. \ref{fig:distance_SMC} and \ref{fig:distance_MW}). In this section, we qualitatively analyse some of the features present in the stellar density map, but it is not in the scope of this paper to analyse all of them in detail. In the next sections we focus only on the bar structure.

As a visual reference for the reader, we show in the left bottom part the face-on density map of the LMC optimal sample \citep{jimenez-arranz23a}. It is evident that two prominent overdensities are the LMC off-centered bar and the spiral arm. We first observe how the LMC spiral arm starts at the end of the bar. Then, if we analyse the spiral arm following a clockwise direction, the spiral arm breaks into two parts: an inner and an outer arm. As mentioned in Section \ref{sec:orbital_history}, those are the observables we are aiming for. For the data, we have no information on the vertical component (edge-on visualisation). That is because of the lack of individual distances we need to assume that all stars lie in the $z'=0$ plane.

First, we qualitatively analyse the presence of galactic bars in our models \citep[e.g.][]{hohl71,noguchi87,berentzen04,besla12,athanassoula13,Yozin2014,cavanagh20,bekki23}. We see that only three models with the \lmc galaxy isolated (leftmost panels) show the presence of a bar: the fiducial model (blue panel), the model with a \lmc unstable disc (light brown panel) and the model with the \lmc with lighter DM halo (orange panel). Otherwise, we observe that the \lmcnospace-\smc interaction triggers the formation of bars in most models (central columns). This is specially evident in the model with the lighter \lmc disc and halo (yellow panel), the model with a \lmc stable disc (light red panel). This interaction does not only trigger the formation of a galactic bar but also can perturb the entire disc in a way that the bar ends up being off-centred (see the fiducial \lmcnospace+\smcnospace+\mw simulation in the face-on top blue right panel), an observable feature of the LMC bar \citep[e.g.][]{Zaritsky2004}.

A second non-axisymmetric feature that is present in all our models is the spiral arms \citep[e.g.][]{williams01,roca-fabrega13,michikosi18,sellwood19,sellwood22}. Most of the isolated simulations (leftmost panels) show flocculent spiral structures if they do not develop a strong bar. When a strong bar is present, the spiral arms are stronger and bisymmetric \citep[see discussion in][]{roca-fabrega13}. In most of the simulations where the interaction with the \smc or \mw systems occurs, we can observe both grand design \citep[e.g.][]{elmegreen82,kendall11} and flocculent spiral arms \citep[e.g.][]{sandage61,elmegreen82,block96,elmegreen03} for different models, regardless a bar is present or not, with a variety of pitch-angles. Nonetheless, in some models we see the formation of a single grand design bisymmetric spiral arm structure also in both cases, when a bar is present and when not. For instance, in the model with lighter \lmc halo (orange right panel) and in the model with lighter \lmc disc and halo (yellow right panel) we see the formation of a high pitch-angle bisymmetric structure. Ring-like structures are also present in some models \citep[as detected in][]{choi2018b}, for example in the three-galaxy simulation of the model with a light \lmc disc but heavy \lmc halo (cyan right panel).

Regarding the \lmc galaxy vertical structure, for the isolated simulations (leftmost panels) we have no noticeable asymmetries in the vertical profile far from a small enhancement in the cases with a bar that underwent or is experiencing a buckling event \citep[left orange panel, e.g.][]{pfenniger91, lokas14}. For the interacting simulations, we observe a variety of vertical asymmetries mostly related to tidal interactions with the \smc and \mw systems, and we see also that the disc is heated up at different degrees depending on the type of interaction.

\subsection{t=0 kinematics}

Given that we possess full information on the position and velocity of each particle, we can show the radial, residual tangential, and vertical velocity maps for the simulated \lmc galaxy disc in the same way that was done for the LMC using \gaia DR3 data \citep[see][]{jimenez-arranz23c}. The \lmc centre-of-mass systemic motion is subtracted in order to obtain the internal velocities.

\begin{figure*}[t!]
    \centering
    \includegraphics[width=0.7\paperwidth]{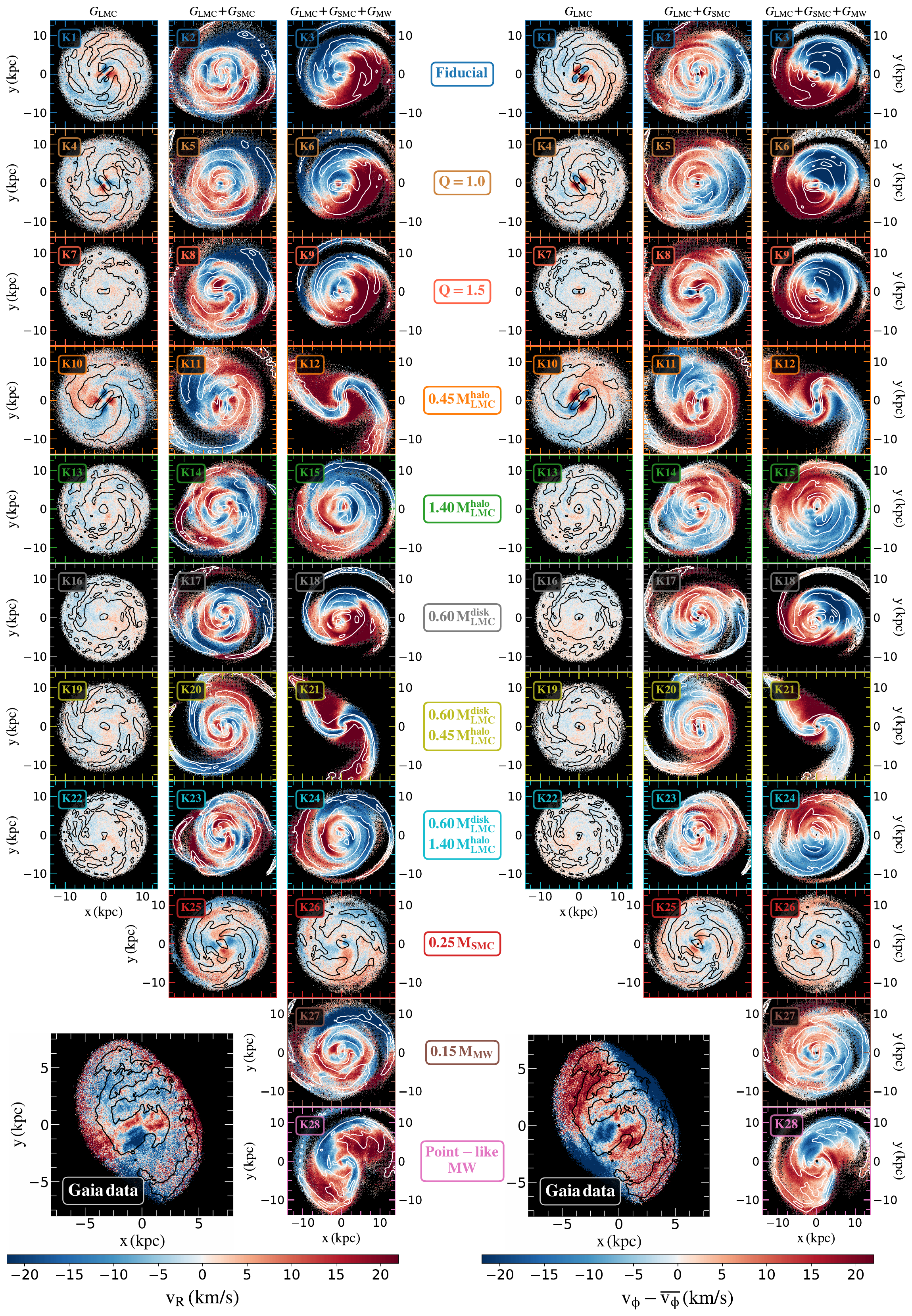}
    \caption{Radial (left part) and residual tangential (right part) velocity maps of the \lmc disc as seen face-on at $t=0$. Each row corresponds to a different set of simulations. For both velocity maps, we have the \lmcnospace, \lmcnospace+\smc and \lmcnospace+\smcnospace+\mw models on the left, centre and right panels, respectively. Black and white contour lines highlight the \lmc overdensities. As reference, the velocity maps of the LMC optimal sample \citep{jimenez-arranz23a} is shown in the bottom left part of each panel.}
    \label{fig:macro_rad_tan_velocity}
\end{figure*}

Figure \ref{fig:macro_rad_tan_velocity} shows the \lmcnospace's disc radial and residual tangential velocity maps as seen face-on (left and right set of columns, respectively) at $t=0$ for all models\footnote{\label{note1}The animation showcasing the evolutionary changes in the radial, residual tangential and vertical distribution throughout the entire simulation is available online.}. When a system is severely perturbed like in the models with \lmcnospace-\smc and \lmcnospace-\smcnospace-\mw interactions, a more detailed inspection is required, albeit we can find systematic changes. The \lmcnospace+\smcnospace+\mw simulation of the fiducial model (top blue right panels), for instance, clearly exhibits bimodality in both radial and tangential velocity. In general, the bimodality becomes more obvious when the \mw system is present (rightmost panels), which is a reflex of the effect of its tides on the \lmc disc.

The kinematic imprint of the bar can also be seen clearly in all models where a bar develops. In the radial and residual tangential velocity maps, the bar produces a quadrupole due to the elliptical orbits of the stars that form it. This is evident in the fiducial and light \lmc halo models where the \lmc is in isolation (blue and orange leftmost panels, respectively). The spiral arms also have a clear signature in the dynamics. For example, the grand design spiral arms seen in the model with lighter \lmc halo (orange right panel) and in the model with lighter \lmc disc and halo (yellow right panel) show clear signs of inwards radial migration (have a negative radial velocity).

Again, we show the radial and tangential velocity maps of the LMC optimal sample \citep{jimenez-arranz23a} in the bottom left part of the corresponding panel as a visual reference for the reader. In both cases the quadrupole pattern that appears in the center of the galaxy is related to the motion of stars in the bar. However, there is an asymmetry clearly apparent along the semi-major axis of the bar. This could be given by the inclination of the bar with respect to the disc \citep{besla12}. Regarding the radial velocity map, there is a negative (inward) motion along the spiral arm overdensity when this is still attached to the bar. After the break, there is no a clear trend. Regarding the residual tangential velocity map, along the spiral arm the residual tangential velocity is in general positive, that is, stars on the spiral arm move faster than the mean motion at the same radius, except for the part of the arm with a density break.

\begin{figure}[t!]
    \centering
    \includegraphics[width=0.85\columnwidth]{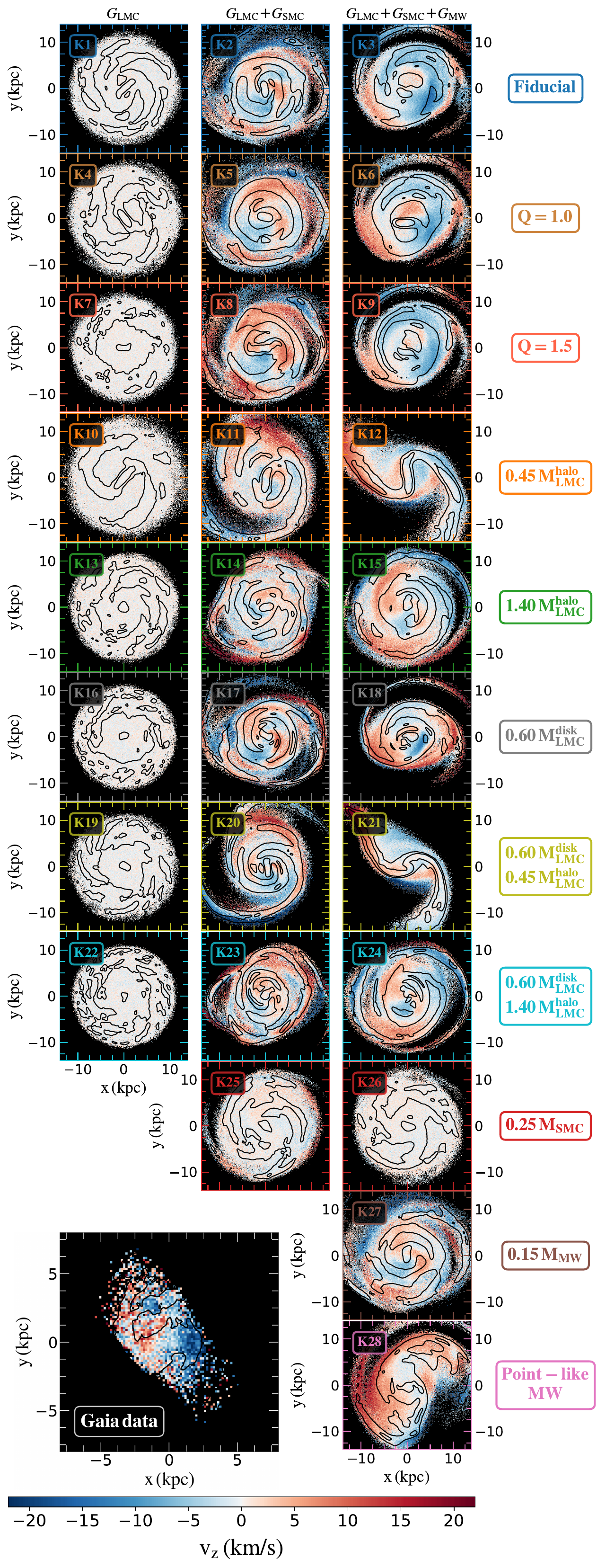}
    \caption{Vertical velocity maps of the \lmc disc as seen face-on at $t=0$. Each row corresponds to a different set of simulations. We have the \lmcnospace, \lmcnospace+\smc and \lmcnospace+\smcnospace+\mw models on the left, centre and right panels, respectively. Black and white contour lines highlight the \lmc overdensities. As reference, the vertical velocity map of the LMC optimal $V_{\text{los}}$ sub-sample \citep{jimenez-arranz23a} is shown in the bottom left part.}
    \label{fig:macro_vert_velocity}
\end{figure}

In Fig. \ref{fig:macro_vert_velocity} we show the \lmc disc vertical velocity maps as seen face-on at $t=0$ for all models\footref{note1}. For the isolated simulations (leftmost panels) we observe no significant vertical velocities, as expected. However, for the interacting simulations (center and rightmost panels), bending modes can be observed \citep[e.g.][]{widrow14,chequers17,chequers18}. That may reflect the vertical structure seen in Fig. \ref{fig:macro_density}. This will be better analysed in a forthcoming paper.

Comparing these maps with the LMC data \citep{jimenez-arranz23a} is difficult because for the data there is not as much information available for the vertical velocity component. This is due to the fact that we need to know $V_{\text{los}}$ in addition to the astrometric information to determine the vertical velocity component, as seen in Section 3 and Appendix A of \citet{jimenez-arranz23a}. Unfortunately, \gaianospace's spectroscopic data is limited to the brightest stars, meaning that our $V_{\text{los}}$ sub-samples only include tens of thousands of stars, while the full samples contain tens of millions sources. The LMC optimal $V_{\text{los}}$ sub-sample's vertical velocity map is shown in the bottom left part of Fig. \ref{fig:macro_vert_velocity}. It exhibits wave-like motion, which may be related to the warp or to the fact that dynamical equilibrium has not yet been reached by the LMC \citep[e.g.][]{Choi2022}. Also, it may indicate that the bar is inclined with respect to the galactic plane.

\section{The \lmc disc scale height}
\label{sec:scale_height}

\begin{figure*}[t!]
    \centering
    \includegraphics[width=0.73\paperwidth]{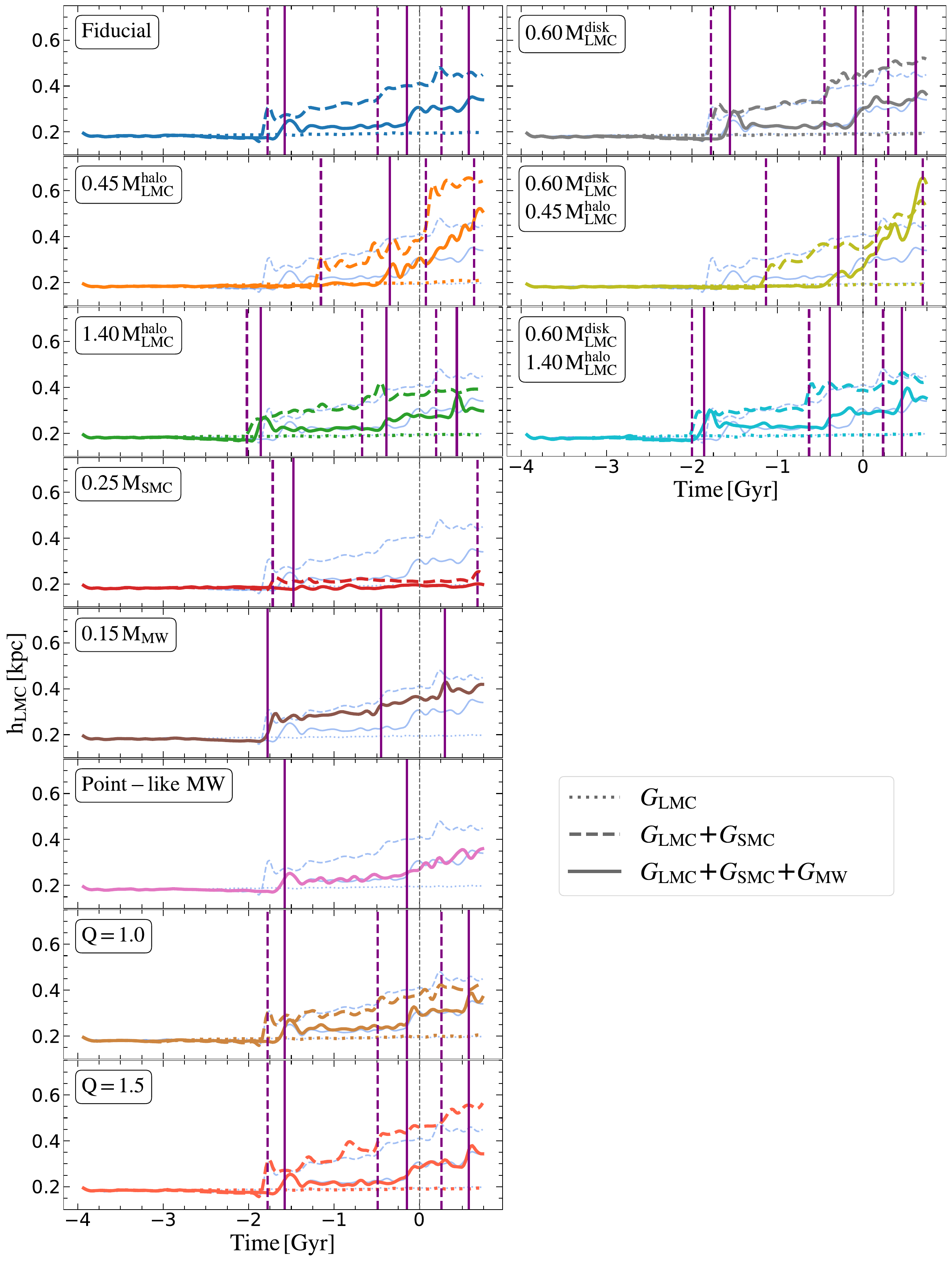}
    \caption{Scale height evolution of the \lmc disc. The dotted, dashed and solid lines show the \lmcnospace, \lmcnospace+\smc and \lmcnospace+\smcnospace+\mw models, respectively. The different panels represent a different set of simulations. The vertical purple solid (dashed) lines correspond to the \lmcnospace-\smc pericenters of the \lmcnospace+\smcnospace+\mw (\lmcnospace+\smcnospace) models. The vertical grey dashed line corresponds to the present time $t=0$. For the sake of comparison, in shadowed blue lines it is plotted the scale height evolution of the \lmc disc for the models of the fiducial set.}
    \label{fig:scale_height}
\end{figure*}

Taking advantage that in simulations we have information of the whole temporal evolution, we can analyse how the individual \lmcnospace-\smc pericenters affect the \lmc disc kinematics. In particular, we show the effect of the pericentric passages on the evolution of the \lmc disc's scale height \citep[e.g.][]{degrijs97,narayan02a,narayan02b}. In Fig. \ref{fig:scale_height} we show the \lmc disc scale height $h_{\text{LMC}}$ as function of time for the different models. In each panel, the dotted, dashed and solid lines represent the \lmcnospace, \lmcnospace+\smc and \lmcnospace+\smcnospace+\mw models, respectively. In each panel, we include the isolated models as a reference point for comparison (dotted lines), and the results from the fiducial models (thin blue shadowed lines). In several models we see that the \lmc disc heating is nearly identical to the fiducial. On the other hand, we see that in models where a strong bar is present, e.g. the low \lmc halo mass model (orange dotted line), or were the \smc system is lighter (red dotted line), the disc heating changes. In particular, for the models where a strong bar is created (low \lmc halo mass and low \lmc disc mass models, orange and light grey lines, respectively) the disc heating jumps up fast after the creation of the bar, while otherwise it remains almost negligible when interacting with a very light \smc system (red lines).

The \lmcnospace-\smc pericenters also correlate well with a sudden increase in disc thickness, and the strength of this change correlates with the pericenter distance, the disc instability, and the merger (\smc system) mass \citep[e.g.][]{quinn93,moetazedian16}. The change in scale height has a peak. After the disc has heated, the thickness slightly decreases. The \lmc disc relaxes to a higher scale height than the original after the \smc initial kick. This occurs following each \smc system pericenter. It is clear that the change in scale height is more impulsive when the \smc system pericenter is closer to the \lmc disc than when it occurs at greater distances, as in the case of the \lmcnospace+\smcnospace+\mw simulation of the fiducial model (blue solid line) vs. the point-like MW (pink solid line) in the second pericenter (notice the pericenter distance in Fig. \ref{fig:distance_SMC}). Similar to the previous point, the simulations without \mw have a thicker \lmc galaxy disc than the simulations with a \mw system, for all models. This might result from the first passages being closer together than they would be in the absence of the \mw system.

If the mass of the \lmc disc is smaller (right panels) the results do not change so much for the fiducial and heavy LMC halo (grey and cyan lines, respectively) but it does for the light \lmc halo (yellow lines).

\section{The \lmc bar}
\label{sec:lmc_bar}

In this section we aim to study the evolution of the \lmc galaxy bar properties in a quantitative way, including the bar strength, length and pattern speed \citep[e.g.][]{tw84,Athanassoula1992,2000debattista,2014sellwood,2019cuomo,2019guo,geron23,2023buttitta} when in isolation and when interacting with the \smc and \mw systems.

To do so, we apply the Dehnen method to the \lmc disc for all available snapshots, for all simulations. The Dehnen method \citep{dehnen23} measures the bar pattern speed $\Omega_p$ and the orientation angle $\phi_b$ of the bar from single snapshots of simulated barred galaxies. The method also determines the bar region, defined by the inner and outer radius $[R_0 ,R_1]$. Hereafter we will refer to $R_1$ as the bar radius or length, as it agrees well with the definition of best estimates for bar lengths in numerical simulations \citep{ghosh-dimatteo23}. For more details about the method see Sect. 2 and Appendix B of \citet{dehnen23}.

\subsection{The \lmc bar strength}

\begin{figure*}[t!]
    \centering
    \includegraphics[width=0.73\paperwidth]{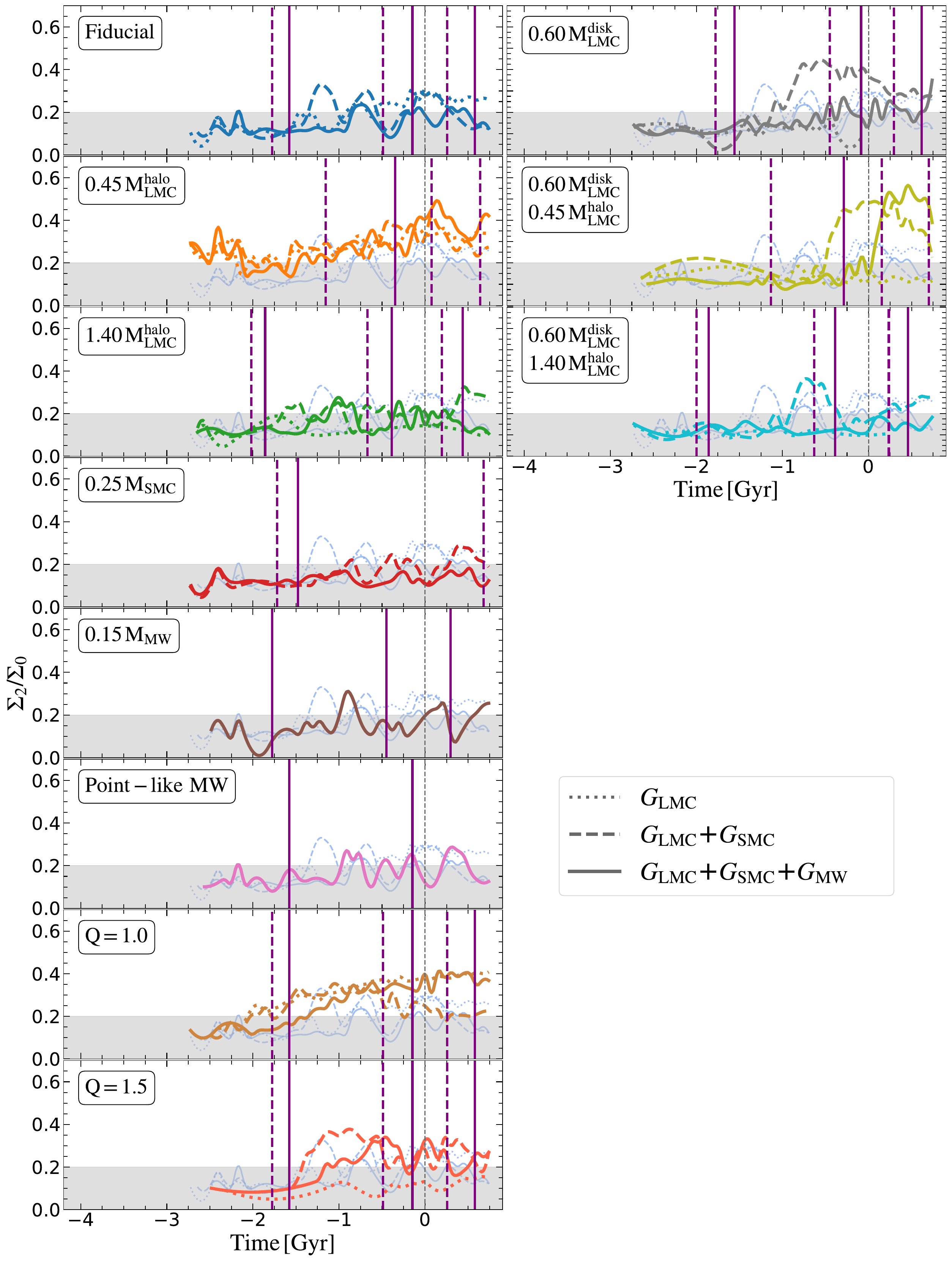}
    \caption{Same as Fig. \ref{fig:scale_height} but for the relative $m = 2$ Fourier amplitude of the \lmc bar region. The grey area corresponds to $\Sigma_2 / \Sigma_0 = 0.2$, which is the threshold used to consider if the \lmc disc has a bar or not. The first 1.25 Gyr are not shown since it is when the \lmc disc is being relaxed, which corresponds to two times the disc rotation.}
    \label{fig:a2}
\end{figure*}

In Fig. \ref{fig:a2}, we show the median relative $m = 2$ Fourier amplitude $\Sigma_2/\Sigma_0$ in the bar region given by the inner and outer radius $[R_0, R_1]$ as function of time. Since the \lmc galaxy disc is relaxing at the beginning of the simulation, we choose not to display the evolution of the disc in the first 1.25~Gyr, which corresponds to two times the disc rotation. Again, in each panel, the dotted, dashed and solid lines represent the \lmcnospace, \lmcnospace+\smc and \lmcnospace+\smcnospace+\mw configurations, respectively. To consider that the disc shows a bar, we impose a threshold of $\Sigma_2/\Sigma_0 > 0.2$ \citep[as in][]{fujii19, bland-hawthorn23}. As qualitatively seen in Fig. \ref{fig:macro_density}, when the \lmc galaxy is in isolation only three models make the disc unstable enough to form a bar; the fiducial model, the \lmc unstable disc model and the \lmc with lighter DM halo model, corresponding to the blue, light brown and orange dotted lines, respectively.

In the fiducial model, if the \smc is present (blue dashed line), we observe how the bar strength is increased $\sim 0.5$ Gyr after the first pericenter and from then, it oscillates. If both \smc and \mw are present (blue solid line), the amplitude of the bar formed is very near the threshold value showing a weak bar. 

Analysing the consequences of having a light or heavy \lmc halo mass on the \lmc bar strength, we observe how reducing the \lmc halo mass (orange lines) makes the disc less stable, allowing it to form a stronger bar in comparison to the fiducial model, for all configurations. The \lmcnospace+\smcnospace+\mw configuration can produce a bar up to $\Sigma_2/\Sigma_0 = 0.4$ for the present time $t=0$. On the contrary, increasing the \lmc halo mass (green lines) makes the disc more stable, making it more difficult to form a bar in comparison to the fiducial model, for all configurations. This is in agreement with simulations in the literature \citep[e.g.][]{roca-fabrega13}.

Regarding the models where the mass of the \lmc disc is smaller (right panels), if we compare all the models of different \lmc halo mass (grey, yellow and cyan lines) with their analogue models with a heavier \lmc disc (blue, orange and green lines, on their left, respectively), we observe that the amplitude of the bar is larger in all cases. This is caused by the fact that, despite being more internally stable, the \lmc disc is more sensitive to external perturbations because the stellar particles are less gravitationally bound.

In the models where the \smc is lighter (red lines), the \mw mass is smaller (dark brown line) and the \mw is considered point-like (pink like), the \lmc shows no bar formation, with differences far from the stochastic variation of the second order Fourier mode with respect to the fiducial model. 

The change of the Toomre parameter $Q$, i.e. the gravitational stability of the stellar disc, has a significant impact on the bar formation for the isolated \lmc models, as expected. The more unstable the disc is (light brown dotted line), the stronger the bar is, independently of whether there is or not interaction with other galactic systems. However, for the \lmcnospace+\smc model, we observe how the bar strength decreases after the second \lmcnospace-\smc pericenter, being the value close to the threshold of $\Sigma_2/\Sigma_0 = 0.2$ at $t=0$. Otherwise, in a more stable disc (light red dotted line), bars are not formed by secular evolution, but we observe how the first \lmcnospace-\smc pericenter boosts the formation of a strong bar on the \lmc galaxy $0.5 - 1$ Gyr after the interaction, for both interacting  models.

\subsection{The \lmc bar length}

\begin{figure*}[t!]
    \centering
    \includegraphics[width=0.7\paperwidth]{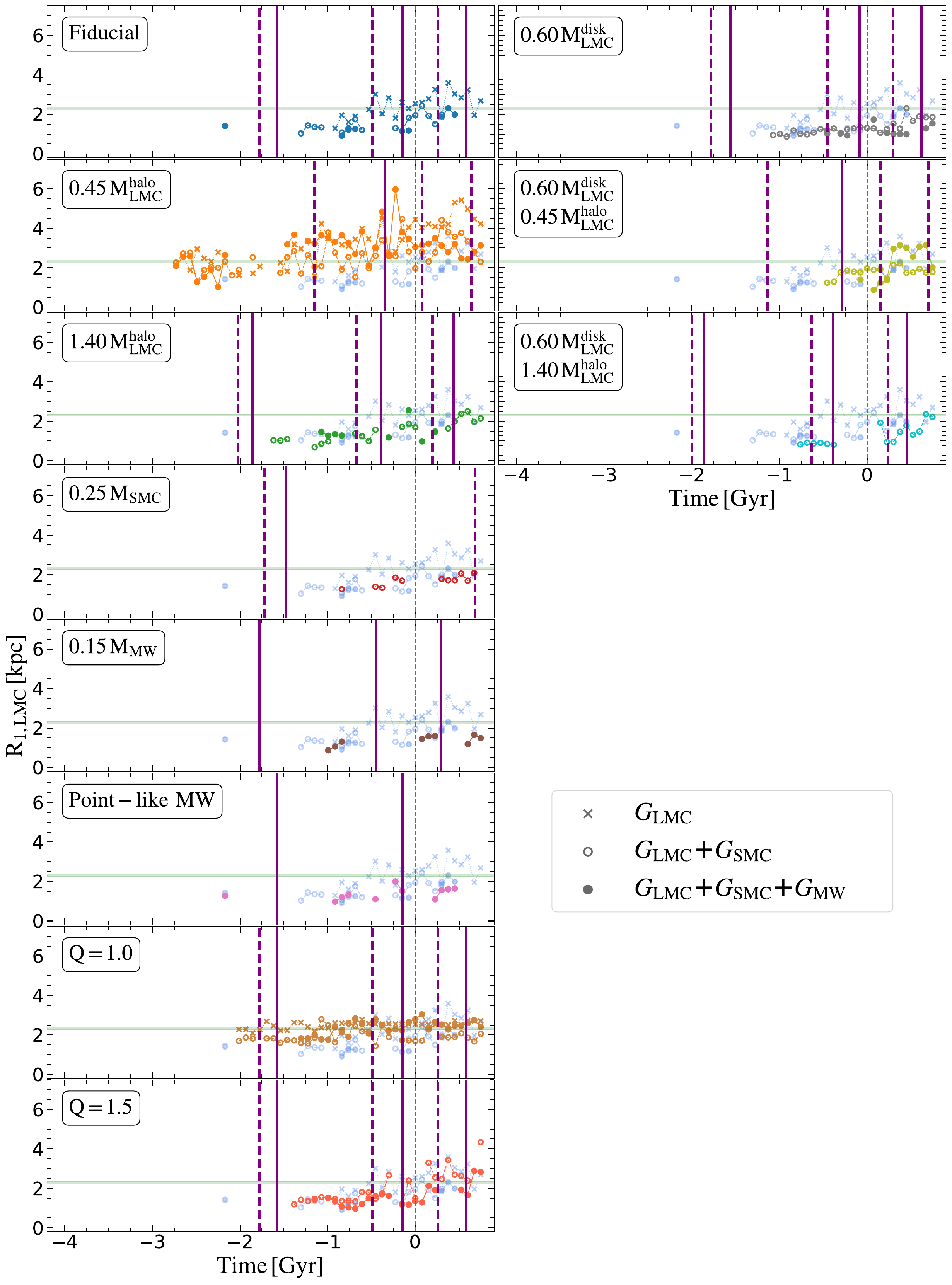}
    \caption{Evolution of the \lmc bar length, given by the outer bar radius $R_1$ from the Dehnen method. The crosses corresponds to the isolated \lmc model, whereas the empty and fill dots show the \lmcnospace+\smc and \lmcnospace+\smcnospace+\mw models, respectively. The different panels represent a different set of simulations. The vertical purple solid (dashed) lines correspond to the MCs pericenters of the \lmcnospace+\smcnospace+\mw (\lmcnospace+\smcnospace) models. The vertical grey dashed line corresponds to the present time $t=0$. The horizontal green area corresponds to the LMC bar length measured in \citet{jimenez-arranz23c}, $R_{\text{1,LMC}} = 2.3$ kpc. For the sake of comparison, in shadowed blue lines it is plotted the evolution of the \lmc bar length for the models of the fiducial set. We only show the obtained value when $\Sigma_2 / \Sigma_0 > 0.2$, which is the threshold used to consider that the \lmc disc has a bar.}
    \label{fig:r1}
\end{figure*}

In Fig. \ref{fig:r1} we show the outer bar region $R_1$ of \lmc given by the Dehnen method as function of time, when the relative $m = 2$ Fourier amplitude $\Sigma_2/\Sigma_0$ is above the 0.2 threshold value. Here and hereafter, the crosses, the empty circles and the filled circles represent the \lmcnospace, \lmcnospace+\smc and \lmcnospace+\smcnospace+\mw configurations, respectively. From the three bars created in isolation (fiducial, low mass \lmc galaxy halo and unstable \lmc disc, represented by the blue, orange and light brown crosses, respectively) the longest is the low mass \lmc galaxy halo model. The horizontal green area corresponds to the LMC bar length measured in \citet{jimenez-arranz23c}, $R_{\text{1,LMC}} = 2.3$ kpc. As happens in the fiducial model, both bars grow over time by a factor $\sim$2 when comparing the end of the simulations with the time when the bar is formed. On the other hand, the model with unstable \lmc disc has a constant bar length $R_{1,\text{LMC}} \sim 2.5$ kpc over time.

We have very complex behaviours on the \lmc bar length for the interacting configurations (empty and filled circles). The shortest bars are of $\lesssim 1$ kpc length and are obtained for interacting models with and without the \mw (see, for example, the grey and green circles). The largest bars are created in the lighter \lmc halo model (orange circles) with a length of $\gtrsim5$ kpc. Whereas for some models the \lmcnospace-\smc pericenters imply a change in the bar length, as for example in the fiducial and heavy \lmc DM halo model (blue and green crosses and circles, respectively), for other models the interaction does not have an impact on the \lmc bar length change, as in the unstable \lmc disc model (light brown crosses and circles).

\subsection{Analysis of the \lmc bar off-centeredness}

\citet{besla12} demonstrated that the off-center stellar bar of the LMC (and its one-armed spiral) can be naturally explained by a recent direct collision with the SMC. Here we analyse how the interaction of the \smc with the \lmc can affect the position of the center of the stellar bar of the \lmcnospace.

\begin{figure*}[t!]
    \centering
    \includegraphics[width=0.73\paperwidth]{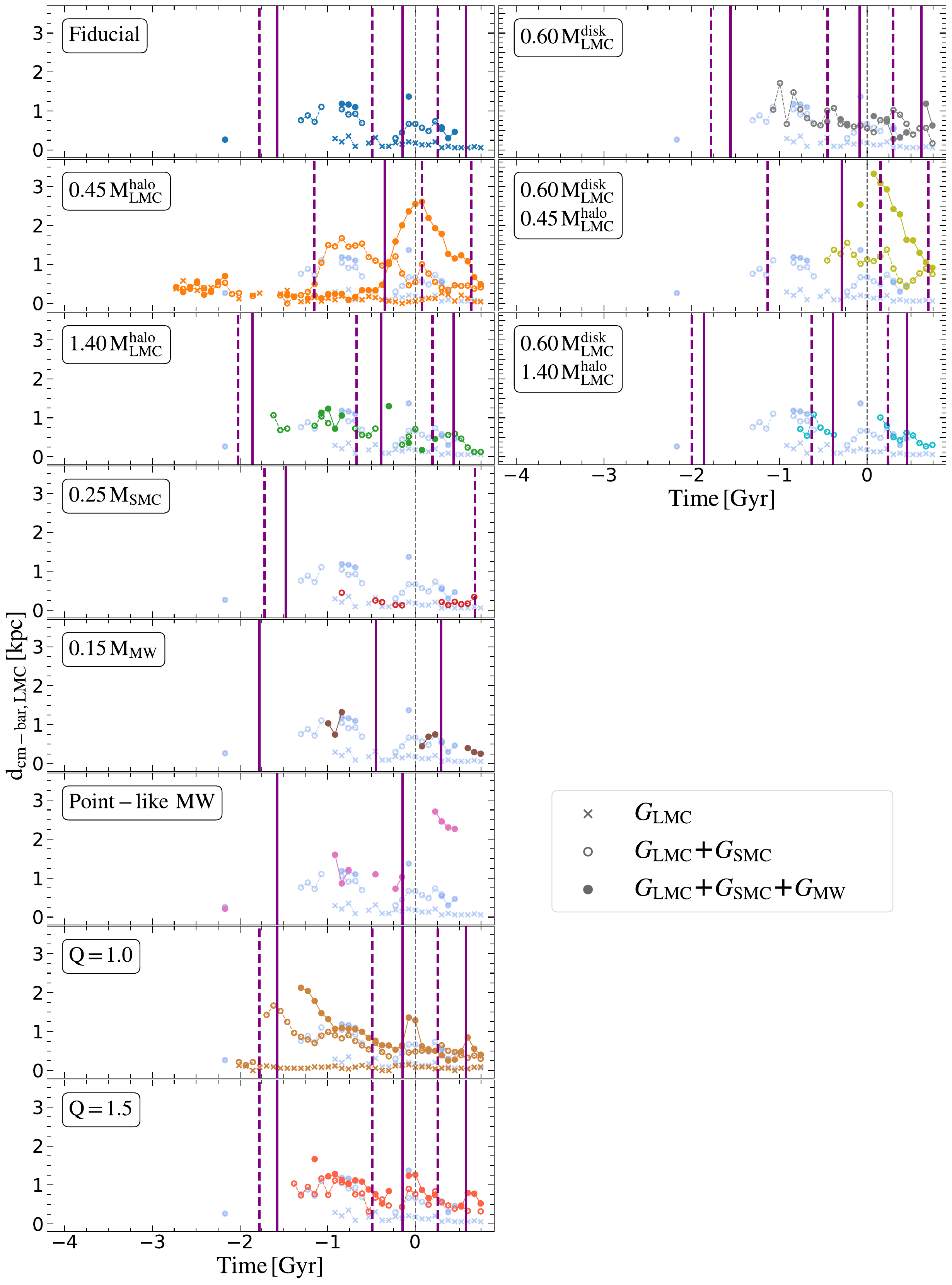}
    \caption{Same as Fig. \ref{fig:r1} but for the \lmc bar off-center. It is given by the distance between the \lmc centre-of-mass and the \lmc bar center obtained using a KDE of 3.0 kpc-bandwidth as explained in Sect. \ref{subsec:analysis_tools}.}
    \label{fig:bar_offcenter}
\end{figure*}

Figure \ref{fig:bar_offcenter} shows the analysis of how off-centered the bar of the \lmc is, determined by the distance between the \lmc centre-of-mass and its bar center (obtained using a KDE of 3.0 kpc-bandwidth, as explained in Sect. \ref{subsec:analysis_tools}), when the relative $m = 2$ Fourier amplitude $\Sigma_2/\Sigma_0$ is above the 0.2 threshold value. The three bars created in isolation (fiducial, low mass \lmc galaxy halo and unstable \lmc disc, represented by the blue, orange and light brown crosses, respectively) share their center with the center-of-mass, as expected, leading to an $\sim0$ kpc bar off-center.

We observe how the \lmcnospace-\smc pericenters in the interacting models produce an increase in the bar off-center. In the full configurations of the light \lmc halo  (orange filled circles) and the light \lmc disc and halo (yellow filled circles) models are where the increased bar off-center is more readily apparent (up to $\sim 2-3$ kpc) $\sim 0.5$ Gyr after the first pericenter and from then, it decreases where the bar tries to be located at the center-of-mass of the host galaxy.  In the majority of simulations we observe an off-center bar at a certain point of the temporal evolution.

\subsection{The \lmc bar pattern speed}

\begin{figure*}[t!]
    \centering
    \includegraphics[width=0.71\paperwidth]{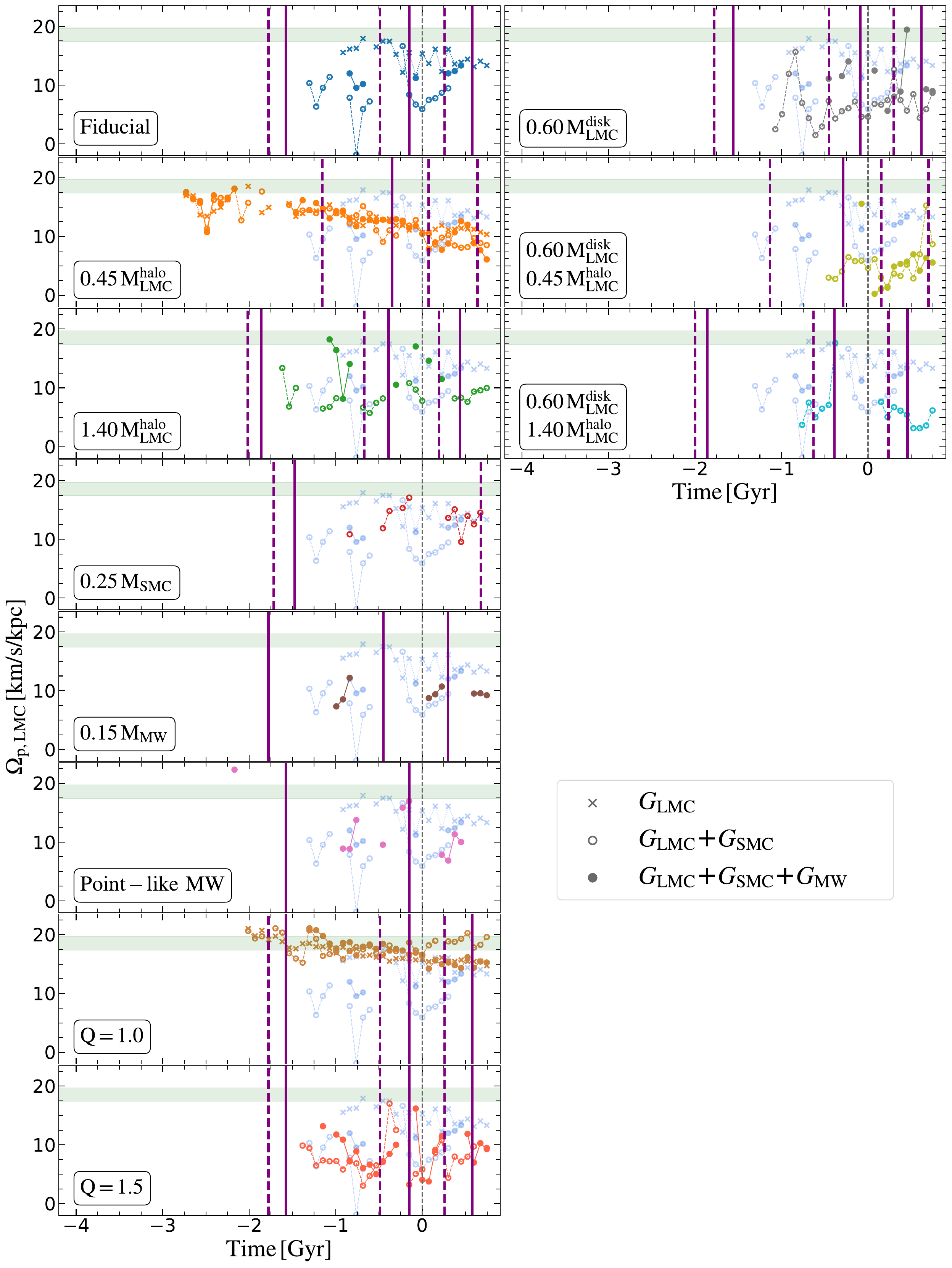}
    \caption{Same as Fig. \ref{fig:r1} but for the \lmc bar pattern speed. The horizontal green area corresponds to the LMC bar pattern speed measured in \citet{jimenez-arranz23c} using the bisymmetric velocity (BV) method \citep{drimmel22}, $\Omega_{\text{p,LMC}} = 18.5^{+1.2}_{-1.1}$ \kmskpc.}
    \label{fig:omegap}
\end{figure*}

Figure \ref{fig:omegap} shows the \lmc bar pattern speed $\Omega_p$ given by the Dehnen method as function of time, when the relative $m = 2$ Fourier amplitude $\Sigma_2/\Sigma_0$ is above the 0.2 threshold value. The horizontal green area corresponds to the LMC bar pattern speed measured in \citet{jimenez-arranz23c} using the bisymmetric velocity (BV) method \citep{drimmel22}, $\Omega_{\text{p,LMC}} = 18.5^{+1.2}_{-1.1}$ \kmskpc. The three bars created in the isolated configuration (fiducial, low mass \lmc galaxy halo and unstable \lmc disc, represented by the blue, orange and light brown crosses, respectively) slow down over time \citep[e.g.][]{2003Athanassoula, 2008Widrow}. In comparison to the fiducial model, the low mass \lmc halo presents a stronger and slower bar, whereas the unstable \lmc disc has a even stronger bar that roughly rotates at the same angular speed.

As in the case of the bar length, we have very complex behaviours for the interacting configurations (empty and filled circles). For the fiducial model, the interacting configurations have bars with smaller pattern speeds than in the isolated configuration. For the interacting configurations corresponding to the two models with more unstable discs (low mass \lmc halo and \lmc disc $Q=1.0$ models, represented by orange and light brown empty and filled circles, respectively) we do not observe significant differences with respect to the decreasing pattern speed shown by the isolated models (crosses). For $t>0$, the \lmcnospace+\smc models show a bar $\sim5$ \kms slower (faster) for the low mass \lmc halo ($Q=1.0$) models in comparison to the isolated model.

\section{Summary and discussion}
\label{sec:discussion}

In this paper we present KRATOS, a comprehensive suite of 28 pure N-body simulations of isolated and interacting LMC-like and SMC-mass galaxies. The 28 simulations are grouped in 11 sets of at most three simulations each containing: 1) a control model with an isolated LMC-like galactic system; 2) a model with both an LMC-like and a SMC-mass system; 3) a model that additionally includes a MW-mass system. For each of the three scenarios, we vary a set of the free parameters of the whole system (details on Sect. \ref{sec:description}). The simulations have a spatial and temporal resolution of 10 pc and 5000 yr, respectively. The minimum mass per particle is $4 \times 10^3$\msun.

The KRATOS suite is devoted to the analysis of the formation of substructures in an LMC-like disc after the interaction with an SMC-mass system and to compare it with the observations \citep[see, for example, the kinematics maps of the LMC using \gaia data on][]{luri20,jimenez-arranz23a}. The majority of the simulations by other authors on the interaction between the LMC, SMC and MW has been done to analyse and recreate the distribution and position of the gas components of these systems, like the Magellanic Stream, Leading Arm, and Magellanic Stream \citep{besla10,besla12,hammer15,pardy18,wang19,tepper-garcia19,lucchini20,lucchini21}.  In the past, only the work by \citet{besla12} extensively explored the effect of this interaction on the internal structures of the LMC such as the off-centered bar and the single spiral arm. It may be possible to learn more about the interactions that took place between these two satellite galaxies and between them and the MW by comprehending the formation process of these LMC morphological features. 

For the fiducial model of the KRATOS suite, we determined that it is more crucial for the \lmc and \smc systems to have completed two pericenters internally than for the distance to the \mw to match observations. Mutual interactions between \lmc and \smc are far more significant than the interaction with the \mw \citep[e.g.][]{devacouleurs-freeman72, besla12}, since tidal effects are smaller at larger distance, even if it is more massive. Although these are not as close to the \mw as one would like, they are nevertheless useful illustrations of the interaction between these two MW neighbours that can be contrasted with observations.

In Table \ref{tabl:kratos_simulations_besla} we compare the initial conditions of the fiducial model of KRATOS suite and the models presented in \citet[][hereafter B12]{besla12}. The B12 models include hydrodynamics whereas the KRATOS suite contains pure N-body simulations. Differences can be found in both the amount of mass of the three galaxies and how it is distributed. For instance, in B12 the LMC is modelled with a disc of gas and a disc of stars surrounded by a DM halo with a Hernquist profile whereas in KRATOS the LMC is modelled as a disc of stars surrounded by a NFW DM halo. A significant difference lies on the modelling of the MW, where B12 models the Galaxy as a static NFW potential whereas in KRATOS simulates the MW as a live NFW DM halo. However, the authors of B12 claim that not considering dynamical friction from the MW halo is expected to have little impact on the orbit in a first passage scenario. Conversely, in both works it is analysed the effect of modelling the LMC and SMC as an isolated binary pair or in interaction with the MW potential.  Finally, in terms of the resolution of both simulations, B12 and KRATOS simulations have of the order of $\sim10^6$ stars particles in the \lmc disc, but there are many more particles on the \lmc DM halo in the KRATOS simulations than in B12 ($10-35$ vs $0.1 \times 10^{6}$). For the \smc, in KRATOS we have less star particles ($0.6$ vs $1 \times 10^{5}$) but a significant larger number of DM particles ($0.01$ vs $1-4.5 \times 10^{6}$). The minimum mass per particle is $2.5 \times 10^3$\msun ${ }$ for B12 and $4 \times 10^3$\msun ${ }$ for KRATOS. The spatial resolution of B12 and KRATOS are 100 and 10 pc, respectively. In \citet{besla12} no explicit information is given on the temporal resolution of B12, while it is of 5000 yr for KRATOS.

\begingroup

\setlength{\tabcolsep}{10pt} 
\renewcommand{\arraystretch}{1.5} 
\begin{table}
\centering
\begin{tabular}{lll}
\hline
\hline
         &  KRATOS  & B12  \\
   & (Pure N-body) & (Hydrodynam.) \\
\hline
\hline
 & [\lmcnospace]  &  \\
\hline
Stellar Distribution   & Disc  & Disc  \\
Scale Length (kpc)   & 2.85  & 1.7  \\
Scale Height (kpc)   & 0.20  & 0.34  \\
Stellar Mass (\msun)   & $5.0 \times 10^9$  & $2.5 \times 10^9$  \\
DM Distribution   & NFW  & Hernquist  \\
DM Concentration    & 9  & 9  \\
DM Mass (\msun)    & $1.8 \times 10^{11}$  & $1.8 \times 10^{11}$  \\
\hline
\hline
 & [\smcnospace] &  \\
\hline
Stellar Distribution   & NFW  & Disc  \\
Stellar Mass (\msun)   & $2.6 \times 10^8$  & $2.6 \times 10^8$  \\
DM Distribution   & NFW  & Hernquist  \\
DM Concentration    & 15  & 15  \\
DM Mass (\msun)    & $1.9 \times 10^{10}$  & $2.1 \times 10^{10}$  \\
\hline
\hline
 & [\mwnospace]  &  \\ 
\hline
DM Distribution   & NFW  & Static NFW  \\
DM Concentration    & 12  & 12  \\
DM Mass (\msun)    & $1.0 \times 10^{12}$  & $1.5 \times 10^{12}$  \\
\hline
\hline
\end{tabular}
\caption{Comparison of the initial conditions of the \lmcnospace, \smc and \mw galaxies for the the fiducial model presented in this work (KRATOS) and the model presented by \citet[][B12]{besla12}.}
\label{tabl:kratos_simulations_besla}
\end{table}

\endgroup

It is possible to compare, accounting for the differences between works, the structure of the LMC stellar disc after the two encounters with the SMC galaxy using both the KRATOS and B12 simulations. As a result of the direct collision between the SMC and the LMC in B12's Model 2, the LMC's bar is off-centred relative to the disc shortly after the galaxies' pericenter. The same outcome is shown in our Fig. \ref{fig:bar_offcenter}, where the off-centeredness of the bar is caused by the tidal interaction between the \lmc and \smcnospace. In B12, the result of this last interaction is a warping of the LMC's bar by approximately $\sim 10^\circ$ - $15^\circ$ with respect to the LMC's disc plane. This analysis is kept for future work on the KRATOS suite. Regarding the formation of a single bar, both B12 (see their Fig. 11, right panel) and KRATOS (our Fig. \ref{fig:macro_density}) works recover this characteristic morphological feature of the LMC. Finally, as our simulations are pure N-body, we are unable to compare with B12 models the gas component of either galaxy.

The KRATOS simulations offer a variety of galaxy formation and evolution, in the sense that \lmc are isolated galaxies, while in \lmcnospace-\smc and \lmcnospace-\smcnospace-\mw  the \lmc is interacting with the \smc or with both \smc and \mwnospace. Under these different scenarios, we can compare the bar fraction in the suite of 28 simulations in KRATOS with that of nearby galaxies, taking into account that they also have different formation histories. In the KRATOS suite, we quantify the number of bars (as used in this work, with $\Sigma_2/\Sigma_0$ amplitude larger than 20\%). We find that in 17 out of the 28 simulations, the disc develops either a weak and transient (less than 1.5 Gyr approx.) or strong bar. Only 11 of the models do not show a clear bar in any moment of the evolution of the galaxy. These numbers agree with the fraction of bars observed in nearby galaxies \citep[e.g.][]{marinova-jogee07,menendez-delmestre07,barazza-jogee-marinova08, sheth08, nair-abraham10b,masters11}, which is about 30\%-60\%. Out of the barred galaxy models, 6 show a weak (below 20\% or very transient in time bar structure), while 11 models show a strong and long-lived bar (more than 1.5Gyr). These strong bars may have different origin: 3 of them lie in the control \lmc models so they develop as internal disc instabilities; 4 of them are formed in models with \lmc and \smc interaction or even with \mw interaction. While 4 bars form in models of interacting galaxies, but the corresponding control \lmc also develops a bar, so both mechanisms could create the bar structure. 

As mentioned above, KRATOS simulations are designed to compare with the internal LMC disc kinematics \citep[e.g.][]{luri20,jimenez-arranz23a}. In \citet{jimenez-arranz23c}, the LMC bar pattern speed is studied using \gaia DR3 data. In this previous work, we use three different methods to evaluate the bar properties: the bar pattern speed is measured through the Tremaine-Weinberg \citep[TW,][]{tw84} method, the Dehnen method (introduced in our Sect. \ref{subsec:analysis_tools}) and a bisymmetric velocity (BV) model \citep{drimmel22}. The work suggests that due to the significant variation with frame orientation, the TW method appears to be unable to determine which global value best represents an LMC bar pattern speed. The Dehnen method gives a pattern speed $\Omega_p= -1.0 \pm 0.5$ \kmskpc, thus corresponding to a bar which barely rotates, slightly counter-rotating with respect to the LMC disc. This method does not take into account a possible strong and counter-rotating $m=1$ disc component, which would balance the bar pattern speed. The BV method recovers a LMC bar corotation radius of $R_c = 4.20 \pm 0.25$ kpc, corresponding to a pattern speed $\Omega_p = 18.5 ^{+1.2}_{-1.1}$ \kmskpc (horizontal green area of Fig. \ref{fig:omegap}). This result is consistent with previous estimates and gives a bar corotation-to-length ratio of $R_c /R_{\text{bar}} = 1.8 \pm 0.1$, which makes the LMC hosting a slow bar, as most interacting bars found in nearby galaxies \citep[e.g.][]{geron23}. The pattern speed of the bars in the KRATOS simulations also fall within this range of $\Omega_p = 10-20$ \kmskpc, suggesting indeed that the LMC has a slow rotating bar.

Regarding the disc scale height computed in the KRATOS simulations and comparing to recent values in the literature, \citet{ripepi22} find, using classical cepheids, that the LMC disc appears ``flared'' and thick, with a disc scale height of $h_{\text{LMC}}=0.97$ kpc. The authors argued that strong tidal interactions with the MW and/or SMC, as well as previous mergers involving now-disrupted LMC satellites, can all be used to explain this feature. Since the present scale height is sensitive to the initial conditions of the \lmc disc, we do not aim to replicate it in our paper. However, we state that interactions indeed lead to a significant increase in the \lmc thickness.

\section{Conclusions}
\label{sec:conclusions}

In this paper we introduce a comprehensive suite of pure N-body and open access simulations named KRATOS with the goal to study the internal kinematics of the LMC disc. \gaia DR3 revealed a complex and rich velocity maps \citep[e.g.][]{luri20,jimenez-arranz23a}. The interpretation of these maps requires a range of simulated LMC models. In this work we generate models with three different configurations: a control \lmc as an isolated LMC galaxy, another one with the \smc only interaction and, finally, the most realistic situation where both the \smc and \mw may interact with the \lmcnospace. We take into account the uncertainties regarding structural parameters and orbital parameters by considering a different set of initial conditions, where we vary one of the parameters at a time (see Table~\ref{tabl:kratos_simulations_all}).

Results shown in this paper from the KRATOS suite can be summarised as:
\begin{itemize}
    \item Regarding the orbital history between the \lmc and the \smcnospace, the effect of not including the \mw makes the pericentric passages between the two galaxies to happen earlier than when the three galaxies are present (see Fig.~\ref{fig:distance_SMC}).
    \item In relation to the orbital history between the \mw and the \lmcnospace, the higher the mass of the \mwnospace, the closer the two galaxies will become, as expected (see Fig.~\ref{fig:distance_MW}).
    \item The KRATOS simulations are suited to explore different \lmc galaxy morphologies, having a large variety of spiral arm shapes, presence of bar, warped discs, etc. as seen in Fig.~\ref{fig:macro_density}.
    \item Different galaxy morphologies also translate into different disc kinematic maps, suitable to perform a first interpretation of the LMC kinematic maps \citep[e.g.][]{luri20, jimenez-arranz23a}, as seen in Figs.~\ref{fig:macro_rad_tan_velocity} and \ref{fig:macro_vert_velocity}.
    \item As Cepheids suggest \citep{ripepi22}, the \lmc disc scale height is increased just after a pericentric passage of the \smc as seen in Fig.~\ref{fig:scale_height} for all models.
    \item Tidal interactions can not only destroy bars (when they are formed), but also create them, as seen in Fig. \ref{fig:a2}.
    \item As demonstrated in \citet{besla12}, we also observe that the off-center stellar bar of the \lmc can be naturally explained by a recent direct interaction with the \smcnospace, as seen in Figs. \ref{fig:macro_density} and \ref{fig:bar_offcenter}.
    \item The \smc pericentric passages do not significantly change the bar length and pattern speed of long-lived bars, as seen in Figs.~\ref{fig:r1} and \ref{fig:omegap}.
    \item As suggested by \citet{geron23}, the longest bars are the ones with a lower pattern speed (see their Fig.~12.). 
\end{itemize}

To sum up, KRATOS suite of N-body simulations are designed to study the internal kinematics of the \lmc disc and to help interpreting the complex maps obtained using \gaia data. The high spatial, temporal and mass resolution used in the simulations are proven to be suitable for such purpose, as the preliminary scientific results presented in this work show. More specific analysis on the LMC-SMC interaction is left for successive papers of the series.

\section*{Acknowledgements}

We thank an anonymous referee for a critical review and constructive suggestions that helped improving the manuscript. This work was (partially) supported by the Spanish MICIN/AEI/10.13039/501100011033 and by "ERDF A way of making Europe" by the “European Union” through grants PID2021-122842OB-C21 and PID2021-125451NA-I00, the European Union «Next Generation EU»/PRTR through grant CNS2022-135232, and the Institute of Cosmos Sciences University of Barcelona (ICCUB, Unidad de Excelencia ’Mar{\'\i}a de Maeztu’) through grant CEX2019-000918-M. OJA acknowledges funding by l'Agència de Gestió d'Ajuts Universitaris i de Recerca (AGAUR) official doctoral program for the development of a R+D+i project under the FI-SDUR grant (2020 FISDU 00011). SRF acknowledges support from the Knut and Alice Wallenberg Foundation, and from the Swedish Research Council (grant 2019-04659). He also acknowledges support from the Spanish Ministry of Science and Innovation through projects PID2020-114581GB-C22 and PID2022-138896NB-C55. MB acknowledges funding from the University of Barcelona’s official doctoral program for the development of a R+D+i project under the PREDOCS-UB grant. PM gratefully acknowledges support from project grants from the Swedish Research Council (Vetenskapr\aa det, Reg: 2017-03721; 2021-04153). LC acknowledges financial support from the Chilean Agencia Nacional de Investigaci\'{o}n y Desarrollo (ANID) through the Fondo Nacional de Desarrollo Cient\'{\i}fico y Tecnol\'{o}gico (FONDECYT) Regular Project 1210992. This work was partially supported by the OCRE awarded project Galactic Research in Cloud Services (Galactic RainCloudS). OCRE receives funding from the European Union’s Horizon 2020 research and innovation programme under grant agreement no. 824079.

\bibliographystyle{aa}
\bibliography{mylmcbib} 

\label{lastpage}

\end{document}